\documentstyle[aps,eqsecnum,preprint,floats,epsf,epsfig]{revtex}
\begin{document}
\def\be{\begin{eqnarray}}
\def\en{\end{eqnarray}}
\def\non{\nonumber}
\def\la{\langle}
\def\ra{\rangle}
\def\nc{N_c^{\rm eff}}
\def\vp{\varepsilon}
\def\a{{\cal A}}
\def\B{{\cal B}}
\def\c{{\cal C}}
\def\d{{\cal D}}
\def\e{{\cal E}}
\def\p{{\cal P}}
\def\t{{\cal T}}
\def\up{\uparrow}
\def\dw{\downarrow}
\def\vma{{_{V-A}}}
\def\vpa{{_{V+A}}}
\def\smp{{_{S-P}}}
\def\spp{{_{S+P}}}
\def\J{{J/\psi}}
\def\ov{\overline}
\def\Lqcd{{\Lambda_{\rm QCD}}}
\def\pr{{\sl Phys. Rev.}~}
\def\prl{{\sl Phys. Rev. Lett.}~}
\def\pl{{\sl Phys. Lett.}~}
\def\np{{\sl Nucl. Phys.}~}
\def\zp{{\sl Z. Phys.}~}
\def\lsim{ {\ \lower-1.2pt\vbox{\hbox{\rlap{$<$}\lower5pt\vbox{\hbox{$\sim$}
}}}\ } }
\def\gsim{ {\ \lower-1.2pt\vbox{\hbox{\rlap{$>$}\lower5pt\vbox{\hbox{$\sim$}
}}}\ } }

\font\el=cmbx10 scaled \magstep2{\obeylines\hfill March, 2002}

\vskip 1.5 cm

\centerline{\large\bf Weak Annihilation and the Effective
Parameters $a_1$ and $a_2$} \centerline{\large\bf in Nonleptonic
$D$ Decays}
\bigskip
\centerline{\bf Hai-Yang Cheng}
\medskip
\centerline{Institute of Physics, Academia Sinica}
\centerline{Taipei, Taiwan 115, Republic of China}
\medskip
\centerline{and}
\medskip
\centerline{C.N. Yang Institute for Theoretical Physics, State
University of New York} \centerline{Stony Brook, New York 11794}
\medskip
\bigskip
\bigskip
\centerline{\bf Abstract}
\bigskip
{\small Based on SU(3) flavor symmetry, many of the quark-graph
amplitudes in two-body nonleptonic decays of charmed mesons can be
extracted from experiment, which enable us to see the relevance
and importance of weak annihilation topologies and to determine
the complex parameters $a_1$ and $a_2$ to test the factorization
approach. It is found that $a_2$ in $D\to \ov K^*\pi$ and $D\to
\ov K\rho$ can be different by a factor of 2, indicating that
nonfactorizable corrections to the latter are far more important
than the former. The relative phase between $a_1$ and $a_2$ is
about $150^\circ$. Weak annihilation topologies induced by nearby
resonances via final-state rescattering  can be described in a
model-independent manner. Although the $W$-exchange contribution
in $D\to PP$ decays is dominated by resonant final-state
interactions (FSIs), its amplitude in $VP$ decays ($V$: vector
meson, $P$: pseudoscalar meson) receives little contributions from
FSIs in the quark-antiquark resonance formation. As a consequence,
the sign flip of the $W$-exchange amplitude in $D\to\ov K^*\pi$
and $\ov K\rho$ decays, which is needed to explain the relatively
real decay amplitudes of $D\to \ov K\rho$, remains unexplained.
SU(3) symmetry is badly broken in some Cabibbo-suppressed modes
and it can be accounted for by the accumulation of some modest
SU(3) violation in individual quark-graph amplitudes.

}

\pagebreak

\section{Introduction}
The hadronic decays of charmed mesons and related physics have
been studied extensively in past 25 years and a lot of progress
has been made. The charm lifetimes, e.g., $\tau(D_s^+)$ and
$\tau(\Xi^+_c)$, the $D^0\!-\!\ov D^0$ mixing and the Dalitz plot
analyses of three-body charm decays are some of the main topics
that are currently being studied. Many new results are expected
soon from the dedicated experiments conducted at CLEO, E791,
FOCUS, SELEX and the $B$ factories BaBar and Belle.

Contrary to the experimental progress, the theoretical advancement
is relatively slow. It is known that the conventional naive
factorization approach fails to describe color-suppressed
(class-II) decay modes. Empirically, it was learned in the 1980s
that if the Fierz-transformed terms characterized by $1/N_c$ are
dropped, the discrepancy between theory and experiment is greatly
improved \cite{Fuk}. This leads to the so-called large-$N_c$
approach for describing hadronic $D$ decays \cite{Buras}.
Theoretically, explicit calculations based on the QCD sum-rule
analysis \cite{BS} indicate that the Fierz terms are indeed
largely compensated by the nonfactorizable corrections. Due to the
success of the $1/N_c$ approach to charmed meson decays, it was
widely believed in the eighties that it applies equally well to
the weak hadronic decays of bottom mesons. However, a
generalization of the large-$N_c$ approach or the sum rule
analysis \cite{Halperin} to hadronic $B$ decays leads to some
predictions in contradiction to experiment, namely, the
destructive interference in the class-III modes $B^-\to
D^{0(*)}\pi^-$ is not borne out by the data. In the heavy quark
limit, nonfactorizable corrections to nonleptonic decays are
calculable due to the suppression of power corrections.
Unfortunately, the charmed quark is not heavy enough to apply the
QCD factorization approach \cite{BBNS} or pQCD in a reliable
manner.

Moreover, the importance of weak annihilation contributions,
namely, $W$-exchange and $W$-annihilation, is controversial. In
practical calculations, it is customary to argue that they are
negligible based on the helicity suppression argument. Although
the observation of $D^0\to\ov K^0\phi$ in the middle 1980s seems
to give the first clean evidence of $W$-exchange, it was claimed
in \cite{Donoghue} that rescattering effects required by unitarity
can produce the same reaction even in the absence of the
$W$-exchange process. Then it was shown in \cite{CC87} that this
rescattering diagram belongs to the generic $W$-exchange topology.
It has been stressed in \cite{CC87} that even in $D\to\ov K\pi$
decays, the $W$-exchange contribution is sizable.

It has been established sometime ago that a least
model-independent analysis of heavy meson decays can be carried
out in the so-called quark-diagram approach. In this diagrammatic
scenario, all two-body nonleptonic weak decays of heavy mesons can
be expressed in terms of six distinct quark diagrams
\cite{Chau,CC86,CC87}:\footnote{Historically, the quark-graph
amplitudes $\t,\,\c,\,\e,\,\a$ are originally denoted by
$\a,\,\B,\,\c,\,\d$, respectively \cite{Chau,CC86,CC87}.} $\t$,
the color-allowed external $W$-emission tree diagram; $\c$, the
color-suppressed internal $W$-emission diagram; $\e$, the
$W$-exchange diagram; $\a$, the $W$-annihilation diagram; ${\cal
P}$, the horizontal $W$-loop diagram ; and ${\cal D}$, the
vertical $W$-loop diagram. (The one-gluon exchange approximation
of the ${\cal P}$ graph is the so-called ``penguin diagram".) It
should be stressed that these quark diagrams are classified
according to the topologies of weak interactions with all strong
interaction effects included and hence they are {\it not} Feynman
graphs. All quark graphs used in this approach are topological and
meant to have all the strong interactions included, i.e. gluon
lines are included in all possible ways. Therefore, topological
graphs can provide information on final-state interactions (FSIs).

Based on SU(3) flavor symmetry, this model-independent analysis
enables us to extract the topological quark-graph amplitudes and
see the relative importance of different underlying decay
mechanisms. The quark-diagram scheme, in addition to be helpful in
organizing the theoretical calculations, can be used to analyze
the experimental data directly. When enough measurements are made
with sufficient accuracy, we can find out the values of each
quark-diagram amplitude from experiment and compare to theoretical
results, especially checking whether there are any final-state
interactions or whether the weak annihilations can be ignored as
often asserted in the literature.

Recently, Rosner \cite{Rosner} has determined the diagrammatic
amplitudes from the measured Cabibbo-allowed two-body $D$ decays.
There are several important observations one can learn from this
analysis. First, the weak annihilation ($W$-exchange or
$W$-annihilation) amplitude is sizable with a large phase relative
to the tree amplitude. Second, the three $D\to \ov K\rho$
amplitudes are observed to be relatively real, in sharp contrast
to $\ov K\pi$ and $\ov K^*\pi$ cases. It was argued in
\cite{Rosner} that the $W$-exchange amplitude has to flip its sign
from $\ov K^*\pi$ to $\ov K\rho$ in order to explain why the three
decay amplitudes of $D\to \ov K\rho$ are in phase with one
another. Third, the color-suppressed amplitude $\c$ has a
nontrivial phase relative to the tree amplitude $\t$. As we shall
see, the appearance of nontrivial relative phases between various
quark-graph amplitudes implies the relevance and importance of
inelastic final-state interactions.

The purpose of this work is twofold: First, we will utilize the
reduced quark-graph amplitudes extracted from the data to
determine the complex parameters $a_1$ and $a_2$ appearing in the
factorization approach. This enables us to test the factorization
hypothesis and see how important the nonfactorizable correction
is. Second, we will study weak annihilations induced from nearby
quark-antiquark intermediate states. This allows us to explore the
effect of inelastic FSIs and see if the sign of the $W$-exchange
topology in Cabibbo-allowed $D\to VP$ decays is governed by nearby
resonances.

The layout of the present paper is as follows. In Sec. II we first
discuss the quark-diagram amplitudes and then extract the
parameters $a_1$ and $a_2$. The diagrammatic amplitudes for
Cabibbo-suppressed decay modes and SU(3) violation are addressed.
The weak annihilation induced from final-state rescattering in
resonance formation is studied in Sec. III. Its implication and
importance for explaining some $D$ decay modes are discussed. Sec.
IV is devoted to exploring the color-suppressed amplitude and its
phase. We then compare the present study with $B$ decays in Sec. V
and give conclusions in Sec. VI.

\section{Diagrammatic approach}

\subsection{Quark-graph amplitudes}
Based on SU(3) flavor symmetry, the quark-graph amplitudes for the
Cabibbo-allowed decays of charmed mesons are listed in Table I.
(For a complete list of the quark-graph amplitudes for Cabibbo
singly and doubly suppressed modes, see \cite{CC87}.)\footnote{For
charm decays involving an SU(3) singlet in the final product,
e.g., $D^0\to \ov K^0\phi,~\ov K^0\omega,~\ov K^0\eta_0$, there
exist additional hairpin diagrams in which a quark-antiquark pair
is created from vacuum to form a color- and flavor-singlet
final-state meson \cite{CC89,Li}. There are four different types
of disconnected hairpin diagrams: $\e_h,\,\a_h,\p_h,\,\d_h$
corresponding to the quark graphs $\e,\,\a,\,\p,\,\d$ (for
details, see \cite{CC89}). Here we will omit the contributions
from the hairpin diagrams, though they seem to play some role in
$D_s^+\to\rho^+\eta'$ \cite{Rosner} and $B\to K\eta'$ decays
\cite{Dighe}.} Note that the selection rule for a vanishing
$D_s^+\to\pi^+\pi^0$ follows from the isospin transformation
properties of the weak Hamiltonian and isospin invariance of
strong interactions and hence it is unaffected by SU(3) breaking
or final-state interactions \cite{Lipkin80}. For final states
involving $\eta$ or $\eta'$ it is more convenient
 to consider the flavor
mixing of $\eta_q$ and $\eta_s$ defined by
 \be
 \eta_q={1\over\sqrt{2}}(u\bar u+d\bar d),\qquad\quad
 \eta_s=s\bar s,
 \en
in analog to the wave functions of $\omega$ and $\phi$ in ideal
mixing. The wave functions of the $\eta$ and $\eta'$ are given by
 \be
 \left(\matrix{ \eta \cr \eta'\cr}\right)=\left(\matrix{ \cos\phi & -\sin\phi \cr
 \sin\phi & \cos\phi\cr}\right)\left(\matrix{\eta_q \cr \eta_s
 \cr}\right),
 \en
where $\phi=\theta+{\rm arctan}\sqrt{2}$, and $\theta$ is the
$\eta\!-\!\eta'$ mixing angle in the octet-singlet basis
 \be
 \left(\matrix{ \eta \cr \eta'\cr}\right)=\left(\matrix{ \cos\theta & -\sin\theta \cr
 \sin\theta & \cos\theta\cr}\right)\left(\matrix{\eta_8 \cr \eta_0
 \cr}\right).
 \en
The $D\to M\eta$ and $M\eta'$ amplitudes have the expressions
 \be \label{etaetap}
 A(D\to M\eta) &=& A(D\to M\eta_8)\cos\phi-A(D\to
 M\eta_0)\sin\phi, \non \\
 A(D\to M\eta') &=& A(D\to M\eta_8)\sin\phi+A(D\to
 M\eta_0)\cos\phi.
 \en

\begin{table}[ht]
\caption{Quark-graph amplitudes for Cabibbo-allowed decays of
charmed mesons. For reduced amplitudes $\t$ and $\c$ in $D\to VP$
decays, the subscript $P$ ($V$) implies a pseudoscalar (vector)
meson which contains the spectator quark of the charmed meson. For
$\e$ and $\a$ amplitudes with the final state $q_1\bar q_2$, the
subscript $P$ ($V$) denotes a pseudoscalar (vector) meson which
contains the antiquark  $\bar q_2$.
 }
\begin{center}
\begin{tabular}{r l|r l|r l  }
 $D^+\to \ov K^0\pi^+$ & $\t+\c$ & $D^+\to \ov K^{*0}\pi^+$ &
$\t_V+\c_P$ &$D^+\to \ov K^0\rho^+$ & $\t_P+\c_V$ \\ \hline$D^0\to
K^-\pi^+$ & $\t+\e$ & $D^0\to K^{*-}\pi^+$ & $\t_V+\e_P$ &
$D^0\to K^-\rho^+$ & $\t_P+\e_V$ \\
$\to \ov K^0\pi^0$ & ${1\over \sqrt{2}}(\c-\e)$ & $\to \ov
K^{*0}\pi^0$ & ${1\over \sqrt{2}}(\c_P-\e_P)$ &
$\to \ov K^0\rho^0$ & ${1\over\sqrt{2}}(\c_V-\e_V)$ \\
 $\to\ov K^0\eta_q$ & ${1\over\sqrt{2}}(\c+\e)$ & $\to\ov
 K^{*0}\eta_q$ & ${1\over \sqrt{2}}(\c_P+\e_P)$ & $\to\ov K^0\omega$
 & ${1\over\sqrt{2}}(\c_V+\e_V)$ \\
 $\to\ov K^0\eta_s$ & $\e$ & $\to\ov
 K^{*0}\eta_s$ & $\e_V$ & $\to\ov K^0\phi$
 & $\e_P$ \\
 \hline
 $D^+_s\to \ov K^0K^+$ & $\c+\a$ & $D_s^+\to \ov K^{*0}K^+$ &
$\c_P+\a_P$ & $D_s^+\to \ov K^0 K^{*+}$ & $\c_V+\a_V$ \\
 $\to \pi^+\pi^0$ & 0 & $\to\rho^+\pi^0$ &
 ${1\over\sqrt{2}}(-\a_P+\a_V)$ & $\to\pi^+\rho^0$ &
 ${1\over\sqrt{2}}(\a_P-\a_V)$ \\
$\to\pi^+\eta_q$ & $\sqrt{2}\a$ & $\to \rho^+\eta_q$ &
${1\over\sqrt{2}}(\a_P+\a_V)$ &
$\to \pi^+\omega$ & ${1\over\sqrt{2}}(\a_P+\a_V)$ \\
$\to\pi^+\eta_s$ & $\t$ & $\to \rho^+\eta_s$ & $\t_P$ & $\to\pi^+\phi$ & $\t_V$ \\
\end{tabular}
\end{center}
\end{table}

Note that the $D\to \ov K\pi$ amplitudes satisfy the isospin
triangle relation
 \be
 A(D^+\to \ov
 K^0\pi^+)=A(D^0\to K^-\pi^+)+\sqrt{2}A(D^0\to \ov K^0\pi^0)
 \en
and likewise for $D\to \ov K^{*}\pi$ and $D\to \ov K\rho$
amplitudes.  Now since all three sides of the isospin triangle are
measured, we are able to determine the relative phases between
various decay amplitudes. From the measured decay rates
\cite{PDG}, we find (only the central values are quoted)
 \be
 &&\delta_{D^0\to\ov K^0\pi^0,D^0\to K^-\pi^+}=30^\circ, \quad
  \delta_{D^+\to\ov K^0\pi^+,D^0\to K^-\pi^+}=80^\circ,
  \non \\
 && \delta_{D^0\to\ov K^{*0}\pi^0,D^0\to
 K^{*-}\pi^+}=20^\circ, \quad
  \delta_{D^+\to\ov K^{*0}\pi^+,D^0\to K^{*-}\pi^+}=97^\circ,
  \non \\
 && \delta_{D^0\to\ov K^0\rho^0,D^0\to K^-\rho^+}\approx 0^\circ, \quad
  \delta_{D^+\to\ov K^0\rho^+,D^0\to K^-\rho^+}\approx 0^\circ,
 \en
where we have used the relation, for example,
 \be
 \cos\delta_{\{\ov K^0\pi^0,K^-\pi^+\}}={\B(D^0\to
 K^-\pi^+)+2\B(D^0\to\ov
 K^0\pi^0)-{\tau(D^0)\over\tau(D^+)}\B(D^+\to\ov K^0\pi^+)\over
 2\sqrt{\B(D^0\to K^-\pi^+)}\sqrt{2\B(D^0\to \ov K^0\pi^0)}}
 \en
to extract the phases. Therefore, the isospin triangle formed by
$\ov K\rho$ amplitudes is dramatically different from the one
constructed by $\ov K\pi$ or $\ov K^*\pi$. This triangle is almost
flat with zero area, indicating that the three $\ov K\rho$
amplitudes are relatively real. This also can be seen from the
isospin analysis:
  \be \label{isospin}
  A(D^0\to K^-\pi^+) &=& \sqrt{2\over 3}A_{1/2}+\sqrt{1\over 3}A_{3/2},   \non  \\
A(D^0\to \ov K^0\pi^0) &=& -\sqrt{1\over 3}A_{1/2} +\sqrt{2\over
3}A_{3/2}, \\ A(D^+\to \ov K^0\pi^+) &=& \sqrt{3}A_{3/2}, \non
 \en
with $A_i=|A_i|e^{i\delta_i}$. It turns out that the isospin phase
difference is consistent with zero for $D\to\ov K\rho$ (see Table
II).

\begin{table}[ht]
\caption{Isospin amplitudes and  phase differences for $D\to\ov
K\pi,\ov K^*\pi,\ov K\rho$ decays.}
\begin{center}
\begin{tabular}{l c c c }
 & $D\to \ov K\pi$ & $D\to\ov K^*\pi$ & $D\to \ov K\rho$ \\
 \hline
 $|A_{1/2}/A_{3/2}|$ & $3.83\pm0.27$  & $5.61\pm 0.35$ & $3.59\pm 0.69$ \\
 $|\delta_{1/2}-\delta_{3/2}|$ & $(90\pm 6)^\circ$ & $(104\pm 13)^\circ$ &
 $<27^\circ$ \\
\end{tabular}
\end{center}
\end{table}

The reduced quark-graph amplitudes $\t,\c,\e$ have been extracted
from Cabibbo-allowed decays by Rosner \cite{Rosner} with the
results:\footnote{The value of the $W$-annihilation amplitude $\a$
is slightly different from that given in \cite{Rosner} as we have
used the $D_s^+$ lifetime:
$\tau(D_s^+)=(0.496^{+0.010}_{-0.009})\times 10^{-12}s$
\cite{PDG}. It seems to us that the phase difference
$|\delta_{AE}|$ quoted in Table IV of \cite{Rosner} is too large
by $10^\circ$.}
 \be \label{PP}
 && \t=2.69\times 10^{-6}\,{\rm GeV}, \qquad\c=(1.96\pm0.14)\,e^{-i152^\circ}\times
 10^{-6}\,{\rm GeV}, \non \\
&&  \e=(1.60\pm0.13)\,e^{i114^\circ}\times 10^{-6}\,{\rm GeV},
\qquad \a=1.10\,e^{-i70^\circ}\times 10^{-6}\,{\rm GeV}
 \en
from $D\to \ov K\pi,\ov K^0\eta,\ov K^0\eta'$ decays, and
 \be \label{VP}
 \t_V=(1.78\pm 0.22)\times 10^{-6}, \quad \c_P=1.48\,e^{-i152^\circ}\times
 10^{-6}, \quad \e_P=(1.39\pm 0.08)\,e^{i96^\circ}\times 10^{-6}
 \en
from $D\to \ov K^*\pi,\,\ov K^0\phi$ and $D_s^+\to\pi^+\phi$.
Without loss of generality, $\t$ has been chosen to be real. The
amplitudes $\t,\c,\e,\a$ have dimensions of energy as they are
related to the decay rate via
 \be
 \Gamma(D\to PP)={p_c\over 8\pi m_D^2}|A|^2,
 \en
with $p_c$ being the c.m. momentum.  In contrast, the reduced
amplitudes $\t_{P,V}, \c_{P,V},\e_{P,V}$ are dimensionless as they
are extracted from the relation
 \be
 \Gamma(D\to VP)={p_c^3\over 8\pi m_V^2}|A|^2.
 \en
Note that our convention for $D\to VP$ amplitudes is the same as
that in \cite{CC87} but  different from those of Rosner
\cite{Rosner}; this allows one to compare the theoretical
calculations of $VP$ amplitudes directly with the quark-graph
amplitudes extracted from experiment. It should be stressed that
the solutions given above are not unique. For example, a small
$\t$ amplitude with the magnitude of $1.1\times 10^{-6}$ GeV is
also allowed. However, it is not favored by the factorization
approach \cite{Rosner,CC87}. The original analysis by Rosner is
based on the $\eta-\eta'$ mixing angle $\theta=-19.5^\circ$ (or
$\phi=35.2^\circ$) from which the $\eta$ and $\eta'$ wave
functions have simple expressions \cite{Chau91}:
 \be
 \eta &=& {1\over \sqrt{3}}(\sqrt{2}\eta_q-\eta_s)={1\over
 \sqrt{3}}(u\bar u+d\bar d-s\bar s),
 \non\\
 \eta'&=& {1\over\sqrt{3}}(\eta_q+\sqrt{2}\eta_s)={1\over
 \sqrt{6}}(u\bar u+d\bar d+2s\bar s).
 \en
From Table I, it is easily seen that $A(D^0\to\ov
K^0\eta)=\c/\sqrt{3}$. Hence, the diagrammatic amplitude $\c$ is
ready to be determined once this mode is measured. A
phenomenological analysis of many different experimental processes
indicates $\theta=-15.4^\circ$ or $\phi=39.3^\circ$ \cite{Kroll}.
However, we find that the above diagrammatical amplitudes
(\ref{PP}) and (\ref{VP}) describe the observed rates well.

In order to extract the quark-graph amplitudes $\t_P,\,\c_V$ and
$\e_V$ from $D\to \ov K\rho,\,\ov K^*\eta$, $\ov K^*\eta',\,\ov
K^0\omega$ decays, some assumptions have to be made. As noted in
passing, the most prominent feature of the $D\to\ov K\rho$ data is
that all their three decay amplitudes are almost in phase with one
another. Therefore, the quark-graph amplitudes have to be aligned
in such a way to render the resultant various decay amplitudes of
$D\to \ov K\rho$ parallel or antiparallel. There exist three
possibilities: (i) $\t_P,\c_V$ and $\e_V$ are relatively real,
(ii) the amplitudes $\t_P$ and $\c_V$ possess a relative phase of
order $-150^\circ$ as in $\ov K\pi$ and $\ov K^*\pi$ cases, but
$\e_V\approx -\e_P$, and (iii) $\e_V$ is close to $\e_P$, but the
relative phase between $\c_V$ and $\t_P$ changes a sign. The first
possibility was first pointed out by Close and Lipkin \cite{CL}.
It turns out that the four data of $D\to\ov K\rho$ and $D^0\to\ov
K^0\omega$ can be fit by setting the three quark-graph amplitudes
to be
 \be \label{set1}
 (i)\quad\qquad \t_P=1.20\times 10^{-6}, \qquad \c_V=0.21\times 10^{-6}, \qquad
 \e_V=1.56\times 10^{-6}.
 \en
As stressed by Close and Lipkin, this fit implies that
$\e_V\gsim\t_P\gg\c_V$ and that the interference in the decay
$D^+\to\ov K^0\rho^+$ is constructive, contrary to the case of
$D^+\to \ov K^0\pi^+$ and $\ov K^{*0}\pi^+$. However, this fit
will be discarded for the following reason. Since $\t_P<\t_V$ and
$\c_V\ll\c_P$, the branching ratios for $D_s^+\to \ov K^0K^{*+}$
and $D_s^+\to\rho^+(\eta,\eta')$ will become quite small.
Neglecting $W$-annihilation amplitudes $\a_V$ and $\a_P$ for the
moment, the fit (\ref{set1}) leads to the predictions
 \be
 \B(D_s^+\to\ov K^0K^{*+})=5.3\times 10^{-4}, \quad
 \B(D_s^+\to\rho^+\eta)=1.1\%, \quad
 \B(D_s^+\to\rho^+\eta')=0.45\%,
 \en
which are too small compared to the corresponding experimental
results \cite{PDG}: $(4.3\pm1.4)\%,~(10.8\pm 3.1)\%,~(10.1\pm
2.8)\%$. As shown below, $\a_P$ and $\a_V$ are constrained by the
measurements of $D_s^+\to\pi^+\rho^0$ and $\pi^+\omega$ and they
are small in magnitude. Within the allowed regions of $\a_V$ and
$\a_P$ constrained by Eqs. (\ref{aP+aV}) and (\ref{aP-aV}), the
predicted branching ratios for aforementioned three modes are
still too small, especially for the $\ov K^0K^{*+}$ decay. For
example, it is found that $\B(D_s^+\to\ov K^0K^{*+})=0.35\%$ for
$\a_V=3.3\times 10^{-7}$. Fit (i) is also unnatural in the sense
that FSIs will generally induce relative phases between various
quark-graph amplitudes.

The second scenario is considered by Rosner \cite{Rosner} based on
the argument that if the $W$-exchange amplitude is dominated by
quark-antiquark intermediate states, then a sign flip of $\e_V$
relative to $\e_P$ will be a consequence of charge-conjugation
invariance. As stressed by Rosner, the presence of large
final-state phases in the $\ov K\rho$ case is masked by the
cancellation between $\c_V$ and $\e_V$. This accidental
cancellation arises in $\ov K\rho$ decays but not in $\ov K\pi$
and $\ov K^*\pi$ decays. In the following we list other two fits
for the quark-graph amplitudes of $\t_P,\c_V$ and $\c_P$:
 \be \label{PV}
(ii)&&\quad \t_P=2.24\times 10^{-6}, \quad
 \c_V=(1.07\pm0.18)e^{-i148^\circ}\times 10^{-6}, \quad
 \e_V=1.20\,e^{-i72^\circ}\times 10^{-6}, \non \\
 (iii)&& \quad \t_P=2.24\times 10^{-6}, \quad
 \c_V=(1.07\pm0.18)e^{i148^\circ}\times 10^{-6}, \quad
 \e_V=1.20\,e^{i72^\circ}\times 10^{-6},
 \en
where fit (ii) was first obtained by \cite{Rosner}. It is easily
seen that although both fits give the same decay rates for $D\to
\ov K\rho$, they yield different predictions for $D^0\to\ov
K^{*0}\eta$. For $\phi=39.3^\circ$, we find $\B(D^0\to\ov
K^{*0}\eta)=1.5\%$ for fit (ii) and $0.78\%$ for fit (iii), while
the experimental branching ratio is $(1.9\pm 0.5)\%$ \cite{PDG}.
Hence, it appears that fit (ii) for quark-graph amplitudes is
preferred. Moreover, as we shall see in Sec. IV, a model
calculation of inelastic final-state rescattering indicates that
the imaginary part of the color-suppressed amplitude $\c_V$ is
negative, in accord with fit (ii). However, it will be shown later
(Sec. III.C) that the $W$-exchange amplitudes $\e_P$ and $\e_V$
are not dominated by resonant FSIs and hence the sign flip of
$\e_V$ from $\e_P$ remains unexplained. Therefore, fit (iii) for
quark-graph amplitudes $\c_V$ and $\e_V$ is not entirely ruled
out. It is worth mentioning that, contrary to fit (i), fit (ii) or
(iii) has relatively large tree and color-suppressed amplitudes.
As a consequence, the interference occurring in $D^+\to\ov
K^0\rho^+$ has to be destructive in order to accommodate the data.

For the $W$-annihilation amplitudes $\a_P$ and $\a_V$, the
measurement of $D_s^+\to\pi^+\omega$ \cite{PDG} leads to
 \be \label{aP+aV}
 |\a_P+\a_V|=(4.5\pm 1.0)\times 10^{-7},
 \en
while the amplitude $|\a_P-\a_V|$ can be extracted from the recent
E791 experiment \cite{E791}
$\Gamma(D_s^+\to\rho^0\pi^+)/\Gamma(D_s^+\to
\pi^+\pi^+\pi^-)=(5.8\pm2.3\pm3.7)\%$, though it does not have
enough statistic significance. The result is
 \be \label{aP-aV}
 |\a_P-\a_V|=(2.0\pm1.2)\times 10^{-7},
 \en
where use of $\B(D_s^+\to\pi^+\pi^+\pi^-)=(1.0\pm 0.4)\%$
\cite{PDG} has been made. It will be shown in Sec. III.B that
$\a_P-\a_V$ receives dominant contributions from resonance-induced
FSIs. Eqs. (\ref{aP+aV}) and (\ref{aP-aV}) suggest that the phase
difference between $\a_P$ and $\a_V$ is less than $90^\circ$ and
that the magnitude of $\a_P$ or $\a_V$ is smaller than $\e_P$ and
$\e_V$, contrary to the $PP$ case. The suppression of
$W$-annihilation will be explained in Sec. III.

Without $W$-annihilation, fit (ii) leads to
  \be
 \B(D_s^+\to\ov K^0K^{*+})=1.4\%, \quad
 \B(D_s^+\to\rho^+\eta)=3.9\%, \quad
 \B(D_s^+\to\rho^+\eta')=1.6\%.
 \en
In the presence of $W$-annihilation contributions, the decay
amplitudes of $D_s^+\to\rho^+\eta^{(')}$ read [see Table I and Eq.
(\ref{etaetap})]
  \be
  A(D_s^+\to\rho^+\eta)&=& \t_P\sin\phi-{1
  \over\sqrt{2}}(\a_P+\a_V)\cos\phi, \non\\
 A(D_s^+\to\rho^+\eta')&=& \t_P\cos\phi+{1
 \over\sqrt{2}}(\a_P+\a_V)\sin\phi.
 \en
It is obvious that $\rho^+\eta$ and $\rho^+\eta'$ cannot be
accommodated simultaneously because if $W$-annihilation
contributes constructively to the former, it will contribute
destructively to the latter, and vice versa. This issue has been
discussed in \cite{Buccella,Ball,CT99,CT01} and it is generally
believed that the large discrepancy between theory and experiment
means that there is an additional contribution to $\rho^+\eta'$
owing to the special character of the $\eta'$. As an illustration,
we find
  \be
 \B(D_s^+\to\ov K^0K^{*+})=2.2\%, \quad
 \B(D_s^+\to\rho^+\eta)=5.4\%, \quad
 \B(D_s^+\to\rho^+\eta')=1.2\%,
 \en
for $\a_P+\a_V=-4.5\times 10^{-7}$ and $\a_V=-3.3\times 10^{-7}$.
The smallness of $\rho^+\eta'$ from tree and $W$-annihilation
contributions may indicate the relevance and importance of the
hairpin diagrams for the $\rho^+\eta'$ decay. For example, an
enhancement mechanism has been suggested in \cite{Ball} that a
$c\bar s$ pair annihilates into a $W^+$ and two gluons, then the
two gluons will hadronize mostly into $\eta'$. The other
possibility is that the gluonic component of the $\eta'$, which
can be identified with the physical state e.g. the gluonium,
couples to two gluons directly.

There are several important observations of the above extracted
reduced quark-graph amplitudes: (i) The $W$-exchange or
$W$-annihilation contribution is in general comparable to the
internal $W$-emission and hence cannot be neglected, as stressed
in \cite{CC87}. The weak annihilation amplitude has a phase of
order $90^\circ$ relative to $\t$ and this is suggestive of the
importance of resonant contributions to weak annihilations in $D$
decays. (ii) The $W$-annihilation $\a$ and $W$-exchange amplitudes
$\e$ have opposite signs. (iii) The color-suppressed internal
$W$-emission amplitude $\c$ has a phase $\sim 150^\circ$ relative
to $\t$ for all Cabibbo-allowed $D$ decays. In the factorization
approach, the relative phase is $180^\circ$. (iv) The $W$-exchange
amplitude in $\ov K^*\pi$ and in $\ov K\rho$ has an opposite sign.
As stressed by Rosner, this sign difference is very crucial to
explain why $D\to \ov K\rho$ amplitudes are relatively real, but
not the case for $D\to \ov K^*\pi$ and $D\to\ov K\pi$ decays.

It was conjectured in \cite{Rosner} that (i) the opposite sign
between $\e_P$ and $\e_V$ arises from the fact that they are
dominated by the quark-antiquark intermediate states which have
equal and opposite couplings to $K^{*-}\pi^+$ and $K^-\rho^+$ by
charge-conjugation invariance, and (ii) the relative phase between
$\c$ and $\t$ comes from inelastic final-state rescattering. We
will come to these issues in Secs. III and IV.

\subsection{Parameters $a_1$ and $a_2$}
In terms of the factorized hadronic matrix elements, one can
define $a_1$ and $a_2$ by
 \be
 \t &=& {G_F\over
 \sqrt{2}}V_{ud}V_{cs}^*a_1(\ov K\pi)\,f_\pi(m_D^2-m_K^2)F_0^{DK}(m_\pi^2),
 \non \\
 \c &=& {G_F\over
 \sqrt{2}}V_{ud}V_{cs}^*a_2(\ov K\pi)\,f_K(m_D^2-m_\pi^2)F_0^{D\pi}(m_K^2),
 \non \\
 \t_V &=& {G_F\over
 \sqrt{2}}V_{ud}V_{cs}^*a_1(\ov K^*\pi)\,2f_\pi m_{K^*}A_0^{DK^*}(m_\pi^2),
 \non \\
 \c_P &=& {G_F\over
 \sqrt{2}}V_{ud}V_{cs}^*a_2(\ov K^*\pi)\,2f_{K^*}m_{K^*}F_1^{D\pi}(m_{K^*}^2),
  \\
 \t_P &=& {G_F\over
 \sqrt{2}}V_{ud}V_{cs}^*a_1(\ov K\rho)\,2f_\rho m_\rho F_1^{DK}(m_\rho^2),
 \non \\
 \c_V &=& {G_F\over
 \sqrt{2}}V_{ud}V_{cs}^*a_2(\ov K\rho)\,2f_K m_\rho
 A_0^{D\rho}(m_K^2),\non
 \en
where we have followed \cite{BSW} for the definition of form
factors. Factorization implies a universal, process-independent
$a_1$ and $a_2$, for example, $a_2(\ov K\rho)=a_2(\ov
K^*\pi)=a_2(\ov K\pi)$.

In order to extract the values of $a_1$ and $a_2$ we consider two
distinct form factors models: the Bauer-Stech-Wirbel (BSW) model
\cite{BSW} and the Melikhov-Stech (MS) model \cite{MS}, both
based on the constituent quark picture. For the $q^2$ dependence,
the BSW model adopts the pole dominance assumption:
 \be
 f(q^2)={f(0)\over (1-q^2/m_*^2)^n},
 \en
with $m_*$ being the pole mass. The original BSW model assumes a
monopole behavior (i.e. $n=1$) for all the form factors. However,
this is not consistent with heavy quark symmetry scaling relations
for heavy-to-light transitions. The modified BSW model takes the
BSW model results for the form factors at zero momentum transfer
but makes a different ansatz for their $q^2$ dependence, namely, a
dipole behavior (i.e. $n=2$) is assumed for the form factors
$F_1,~A_0,~A_2,~V$, motivated by heavy quark symmetry, and a
monopole dependence for $F_0,A_1$. The experimental value of
$F_0^{DK}(0)$ is about 0.76 \cite{CLEO93}, but it is far less
certain for $F_0^{D\pi}(0)$. A sum rule analysis \cite{LCSR} and
in particular a recent lattice calculation \cite{Abada} all give
$F_0^{DK}(0)/F_0^{D\pi}(0)\approx 1.20$\,, in agreement with the
results of the BSW and MS models (see Table III).

\begin{table}[ht]
\caption{Form factors in BSW and MS models.}
\begin{center}
\begin{tabular}{l c c c c c c }
 & $F_0^{DK}(m_\pi^2)$ & $F_0^{D\pi}(m_K^2)$ & $F_1^{DK}(m_\rho^2)$
 & $F_1^{D\pi}(m_{K^*}^2)$ &
 $A_0^{DK^*}(m_\pi^2)$ & $A_0^{D\rho}(m_K^2)$ \\ \hline
BSW & 0.76 & 0.72 & 1.01 & 1.07 & 0.74 & 0.77 \\
MS  & 0.78 & 0.71 & 0.93 & 0.91 & 0.76 & 0.73 \\
\end{tabular}
\end{center}
\end{table}

The values of $a_1$ and $a_2$ and their ratio are listed in Table
IV. We see that the ratio of $a_2/a_1$ is channel dependent,
especially its magnitude in $\ov K^*\pi$ and $\ov K\rho$ decays
can be different by a factor of 2. However, its phase of order
$150^\circ$ is essentially process independent. To see the
sensitivity of $a_2/a_1$ to the $W$-exchange contribution, we set
$\e=0$ and determine $a_2/a_1$ from the isospin analysis of $D\to
\ov K^{(*)}\pi(\rho)$ decays. For example, for $D\to\ov K\pi$
decays we have \cite{cheng01}
 \be
 \left.{a_2\over a_1}\right|_{D\to\ov K\pi}
= {2-\sqrt{2}\left.{A_{1/2}\over A_{3/2}}\right|_{D\to\ov
K\pi}\over
 1+\sqrt{2}\left.{A_{1/2}\over A_{3/2}}\right|_{D\to\ov K\pi} }
 \,{f_\pi\over f_K}\,{m_D^2-m_K^2\over
 m_D^2-m_\pi^2}\,{F_0^{DK}(m_\pi^2)\over
 F_0^{D\pi}(m_K^2)}.
 \en
The results are shown in the last two rows of Table IV. It is
interesting to see that for $\ov K\pi$ and $\ov K^*\pi$ decays,
the phase of $a_2/a_1$ is about the same as before but the
magnitude differs slightly, whereas for the $\ov K\rho$ system,
the magnitude is close to the realistic one but the phase is
different. This is understandable because the three $\ov K\rho$
amplitudes are in phase with one another. Hence, $a_2/a_1$ is real
in the absence of the weak annihilation.

\begin{table}[ht]
\caption{The parameters $a_1$ and $a_2$ extracted using the BSW
model (fist entry) and the MS model (second entry) for form
factors. Only the central values for the magnitude and the phase
angle are quoted. To see the sensitivity of $a_2/a_1$ to the
$W$-exchange contribution, its value in the absence of $\e$ is
also shown in last two rows.}
\begin{center}
\begin{tabular}{l c c c }
 & $D\to \ov K\pi$ & $D\to\ov K^*\pi$ & $D\to \ov K\rho$ \\
 \hline
 $|a_1|$ & 1.02 & 1.23 & 0.92 \\
       & 1.05 & 1.28 & 0.85 \\
 $|a_2|$ & $0.63$ &  $0.53$ & 0.76 \\
  & 0.62 &  0.45 & 0.72 \\
 $a_2/ a_1$ & $0.62\,{\rm exp}(-i152^\circ)$ &  $0.43\,{\rm
 exp}(-i152^\circ)$ & $0.82{\,\rm exp}(-i148^\circ)$ \\
  & $0.60\,{\rm exp}(-i152^\circ)$ &  $0.35\,{\rm
 exp}(-i152^\circ)$ & $0.85\,{\rm exp}(-i148^\circ)$ \\
 \hline
 $a_2/ a_1$ (with $\e=0$) & $0.88\,{\rm exp}(-i149^\circ)$ &  $0.56\,{\rm
 exp}(-i160^\circ)$ & $-0.87$ \\
  & $0.86\,{\rm exp}(-i149^\circ)$ &  $0.46\,{\rm
 exp}(-i160^\circ)$ & $-0.90$ \\
\end{tabular}
\end{center}
\end{table}

What is the relation between the coefficients $a_i$ and the Wilson
coefficients in the effective Hamiltonian approach ? Under the
naive factorization hypothesis, one has
 \be \label{nf}
a_1(\mu)=c_1(\mu)+{1\over N_c}c_2(\mu), \qquad \quad
a_2(\mu)=c_2(\mu)+{1\over N_c}c_1(\mu),
 \en
for decay amplitudes induced by current-current operators
$O_{1,2}(\mu)$, where $c_{1,2}(\mu)$ are the corresponding Wilson
coefficients and $N_c$ is the number of colors. In the absence of
QCD corrections, $c_1=1$ and $c_2=0$, and hence class-II modes
governed by $a_2=1/N_c$ are obviously ``color-suppressed".
However, this naive factorization approach encounters two
principal difficulties: (i) the coefficients $a_i$ given by Eq.
(\ref{nf}) are renormalization scale and $\gamma_5$-scheme
dependent, and (ii) it fails to describe the color-suppressed
class-II decay modes due to the smallness of $a_2$. Therefore, it
is necessary to take into account nonfactorizable corrections:
 \begin{eqnarray}  \label{a12}
a_1= c_1(\mu) + c_2(\mu) \left({1\over N_c} +\chi_1(\mu)\right)\,,
\qquad \quad a_2 = c_2(\mu) + c_1(\mu)\left({1\over N_c} +
\chi_2(\mu)\right)\,,
\end{eqnarray}
where nonfactorizable terms are characterized by the parameters
$\chi_i$, which receive corrections including vertex corrections,
hard spectator interactions involving the spectator quark of the
heavy meson, and  FSI effects from inelastic rescattering,
resonance effects, $\cdots$, etc. The nonfactorizable terms
$\chi_i(\mu)$ will compensate the scale and scheme dependence of
Wilson coefficients to render $a_i$ physical. Using the leading
order Wilson coefficients $c_1(\bar m_c)=1.274$ and $c_2(\bar
m_c)=-0.529$ \cite{Buras96} for $\Lambda_{\ov{\rm MS}}=215$ MeV,
where $\bar m_c(m_c)\approx 1.3$ GeV, it is clear that $a_2$ is
rather sensitive to $\chi_2$ and that the nonfactorizable
correction to $a_2$ in the $\ov K\rho$ system is far more
important than that in $\ov K^*\pi$ decays.

Empirically, it was found that that the discrepancy between theory
and experiment for charm decays is greatly improved if Fierz
transformed terms in (\ref{nf}) are dropped \cite{Fuk}. It has
been argued that this empirical observation is justified in the
so-called large-$N_c$ approach in which a rule of discarding
subleading $1/N_c$ terms can be formulated \cite{Buras}. This
amounts to having universal nonfactorizable terms
$\chi_1=\chi_2=-1/N_c$ in Eq. (\ref{a12}) and hence
 \be
 a_1=c_1(\bar m_c)\approx 1.27\,, \qquad\qquad a_2=c_2(\bar m_c)\approx -0.53\,.
 \en
This corresponds to a relative phase of $180^\circ$. From Table IV
we see that the above values of $a_1$ and $a_2$ give a good
description of the $D\to \ov K^*\pi$ decays and differ not too
much from those values for $\ov K\pi$ and $\ov K\rho$ systems.
Hence, $a_1$ and $a_2$ in the large-$N_c$ approach can be
considered as the benchmarked values. In the heavy quark limit,
nonfactorizable terms $\chi_i$ are calculable due to the
suppression of power corrections provided that the emitted meson
is light while the recoiled meson can be either light or heavy
\cite{BBNS}. In the QCD factorization approach, $\chi_i$ are found
to be positive for $B$ decays \cite{BBNS}. This means that the
empiric large-$N_c$ approach cannot be generalized to the $B$
system. For charm decays, the charmed quark is not heavy enough to
apply the QCD factorization approach or pQCD in a reliable manner.
To our knowledge, the sum-rule approach is more suitable for
studying the nonfactorized effects in hadronic $D$ decay
\cite{BS}.

\subsection{Cabibbo-suppressed modes and SU(3) breaking}
So far the diagrammatic amplitudes are determined for
Cabibbo-allowed $D$ decays. When generalized to Cabibbo-suppressed
modes, there exist some sizable SU(3) breaking effects which
cannot be ignored. Table V shows the predicted branching ratios
(see column 3 denoted by $\B_{\rm theory1}$) for some of the
Cabibbo-suppressed modes using the reduced amplitudes determined
from Cabibbo-allowed decays. By comparing with experiment, we see
some large discrepancies. To be specific, we consider the ratios:
 \be
 R_1=2\left|{V_{cs}\over
 V_{cd}}\right|^2\,{\Gamma(D^+\to\pi^0\pi^+)\over\Gamma(D^+\to \ov
 K^0\pi^+)}, \qquad\qquad  R_2={\Gamma(D^0\to
 K^+K^-)\over\Gamma(D^0\to\pi^+\pi^-)}.
 \en
In the SU(3) limit, $R_1=R_2=1$, while the experimental
measurements $R_1=3.39\pm 0.70$ and $R_2=2.80\pm 0.20$ \cite{PDG}
show a large deviation from SU(3) flavor symmetry.

\begin{table}[ht]
\caption{Predicted branching ratios of some Cabibbo-suppressed $D$
decays without and with SU(3) violation (denoted by the subscripts
``theory1" and ``theory2", respectively) and comparison with
experiment. The amplitude $\e_q$ in $D^0\to K^0\ov K^0$ decay
denotes the $q\bar q$-popping $W$-exchange amplitude.}
\begin{center}
\begin{tabular}{r c c c l }
Decay mode & Amplitude & $\B_{\rm theory1}$ & $\B_{\rm theory2}$ & $\B_{\rm expt}$ \cite{PDG} \\
\hline
 $D^+\to \pi^+\pi^0$ & ${1\over\sqrt{2}}{V_{cd}\over V_{cs}}(\t+\c)_{\pi\pi}$ &
 $7.6\times 10^{-4}$ & $2.4\times 10^{-3}$ & $(2.5\pm0.7)\times 10^{-3}$ \\
 $D^0\to\pi^+\pi^-$ & ${V_{cd}\over V_{cs}}(\t+\e)_{\pi\pi}$ & $2.1\times
 10^{-3}$ & $1.7\times 10^{-3}$ & $(1.52\pm 0.09)\times 10^{-3}$ \\
 $\to \pi^0\pi^0$ & ${\sqrt{2}\over 2}{V_{cd}\over V_{cs}}(\c-\e)_{\pi\pi}$ & $1.2\times
 10^{-3}$ & $0.9\times 10^{-3}$ & $(8.4\pm 2.2)\times 10^{-4}$ \\ \hline
 $D^+\to K^+\ov K^0$ & ${V_{us}\over V_{ud}}(\t-\a)_{KK}$ &
 $4.7\times 10^{-3}$ & $7.4\times 10^{-3}$ & $(7.4\pm1.0)\times 10^{-3}$ \\
 $D^0\to K^+K^-$ & ${V_{us}\over V_{ud}}(\t+\e)_{KK}$ & $1.8\times
 10^{-3}$ & $4.2\times 10^{-3}$ & $(4.25\pm 0.16)\times 10^{-3}$ \\
 $\to K^0\ov K^0$ & ${V_{us}\over V_{ud}}(\e_s-\e_d)_{KK}$ & -- & -- & $(6.5\pm 1.8)\times 10^{-4}$ \\
\end{tabular}
\end{center}
\end{table}

As first stressed in \cite{CC94}, model predictions are very
difficult to accommodate the measured value of $R_1$. It was
originally argued in the same reference that large SU(3) violation
manifested in $R_1$ can be accounted for by the accumulations of
several small SU(3) breaking effects, provided that
$F_0^{D\pi}(0)>F_0^{DK}(0)$. However, smaller $F_0^{D\pi}(0)$ is
preferred on theoretical grounds as discussed before. To
accommodate the data of $\pi^+\pi^0$, it is clear that one needs
$\t_{\pi\pi}>\t$ and $|\c_{\pi\pi}|<|\c|$. Allowing modest SU(3)
violation in the individual quark-graph amplitudes,
 \be \label{pipiamp}
&& \t_{\pi\pi}=1.25\,\t, \qquad
\c_{\pi\pi}=1.6\,e^{-i140^\circ}\times
 10^{-6}\,{\rm GeV}, \non \\
 && \e_{\pi\pi}=1.9\,e^{i140^\circ}\times 10^{-6}\,{\rm GeV},
 \en
the data of $D\to\pi\pi$ are well accounted for (see Table V).
However, we would like to stress that we do not claim that Eq.
(\ref{pipiamp}) is {\it the} solution for Cabibbo-suppressed $D\to
\pi\pi$ decays; we simply wish to illustrate that {\it modest}
SU(3) violation in each quark-graph amplitudes can lead to a {\it
large} SU(3)-breaking effect for $R_1$. In the factorization
approach, however, it is difficult to understand why
$\t_{\pi\pi}>\t$.

The ratio of $K^+K^-$ to $\pi^+\pi^-$ is a long-standing puzzle.
The conventional factorization approach leads to $R_2\approx 1$
when weak annihilation contributions are neglected. In the
diagrammatical approach, it is found that $R_2$ can be
accommodated by allowing SU(3) violation in the tree and
$W$-exchange amplitudes:
 \be
 \t_{KK}=1.25\t, \quad \e_{KK}=1.7\,e^{i90^\circ}\times 10^{-6}\,{\rm
 GeV},\quad \a_{KK}=\a.
 \en

It is known that in the limit of SU(3) symmetry, $D^0\to K^0\ov
K^0$ vanishes. This decay receives contributions from inelastic
final-state scattering in analog to Fig. 1(a) and it has been
discussed recently in \cite{Dai,Eeg}.

\section{Weak annihilations and resonant final-state
interactions}
 We learned from Sec. II some important information about the weak
annihilation topologies $\e$ and $\a$: (i) Their contributions are
in general comparable to the internal $W$-emission topological
amplitude and they have a phase of order $90^\circ$ relative to
$\t$ with an opposite sign between $\e$ and $\a$. (ii) A sign flip
of the $W$-exchange amplitude may occur in $\ov K^*\pi$ and $\ov
K\rho$ decays and this is very crucial for explaining why $D\to
\ov K\rho$ amplitudes are in phase with one another, but not the
case for $D\to \ov K^*\pi$ and $D\to\ov K\pi$ decays. The purpose
of this section is to explore these two features.

Under the factorization hypothesis, the factorizable $W$-exchange
and $W$-annihilation amplitudes are proportional to $a_2$ and
$a_1$, respectively. They are suppressed due to the smallness of
the form factor $F_0^{0\to \ov K\pi}(m_D^2)$ at large $q^2=m_D^2$.
This corresponds to the so-called helicity suppression. At first
glance, it appears that the factorizable weak annihilation
amplitudes are too small to be consistent with experiment at all.
However, in the diagrammatic approach here, the topological
amplitudes $\c,~\e,~\a$ can receive contributions from the tree
amplitude $\t$ via final-state rescattering, as illustrated in
Fig. 1 for $D^0\to\ov K^0\pi^0$ decay: Fig. 1(a) has the same
topology as $W$-exchange,\footnote{It is also pointed out by Close
and Lipkin \cite{CL} that the prominent weak annihilation may be
largely due to final-state resonance scattering.} while 1(b)
mimics the internal $W$-emission amplitude $\c$. Therefore, even
if the short-distance $W$-exchange vanishes, a long-distance
$W$-exchange can be induced via inelastic FSIs \cite{Zen,Neubert}.
Historically, it was first pointed out in \cite{Donoghue} that
rescattering effects required by unitarity can produce the
reaction $D^0\to\ov K^0\phi$, for example, even in the absence of
$W$-exchange diagram. Then it was shown in \cite{CC87} that this
rescattering diagram belongs to the generic $W$-exchange topology.

\begin{figure}[t]
\vspace{-4cm}
  \psfig{figure=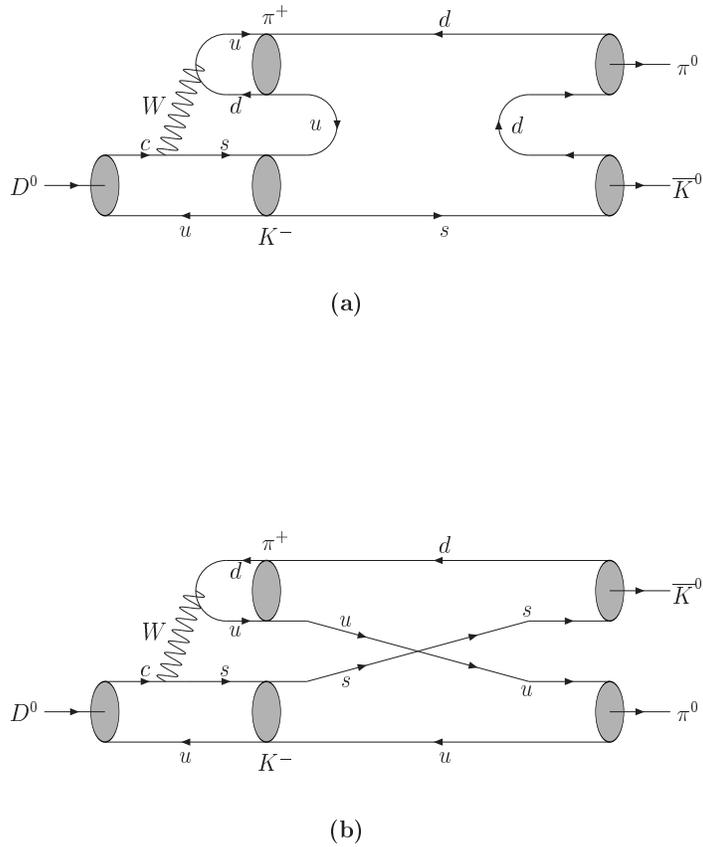,width=16cm}
\vspace{-7.5cm}
    \caption[]{\small Contributions to $D^0\to \ov K^0\pi^0$ from
    the color-allowed weak decay $D^0\to K^-\pi^+$ followed by a
    resonant-like rescattering (a) and quark exchange (b). While (a)
    has the same topology as the $W$-exchange graph, (b) mimics
    the color-suppressed internal $W$-emission graph.}
\end{figure}

There are several different forms of FSIs: elastic scattering and
inelastic scattering such as quark exchange, resonance
formation,$\cdots$, etc. As emphasized in \cite{Zen}, the
resonance formation of FSI via $q\bar q$ resonances is probably
the most important one. Indeed, there are two indications about
the importance of resonant FSIs for weak annihilation topologies:
First, the sizable magnitude of $\e$ and $\a$ and their large
phases are suggestive of nearby resonance effects. Second, an
abundant spectrum of resonances is known to exist at energies
close to the mass of the charmed meson.

Since FSIs are nonperturbative in nature, in principle it is
notoriously difficult to calculate their effects. It is customary
to evaluate the long-distance $W$-exchange contribution, Fig.
1(a), at the hadron level manifested as Fig. 2
\cite{Dai,aaoud,Ablikim,Gronau}. Take $D^0\to \ov K^0\pi^0$ as an
illustration. Fig. 2(a) shows the resonant amplitude coming from
$D^0\to K^-\pi^+$ followed by a $s$-channel $J^P=0^+$ particle
exchange with the quark content $(s\bar d)$, for example,
$K^*_0(1950)$, which couples to $\ov K^0\pi^0$ and $K^-\pi^+$.
Fig. 2(b) corresponds to the $t$-channel contribution with
one-particle exchange. As discussed before, it is expected that
the long-distance $W$-exchange is dominated by resonant FSIs as
shown in Fig. 2(a). However, a direct calculation of this diagram
is subject to many theoretical uncertainties. For example, the
coupling of the resonance to $\ov K\pi$ states is unknown and the
off-shell effects in the chiral loop should be properly addressed
\cite{aaoud}. Nevertheless, as pointed out in \cite{Zen,Weinberg},
most of the properties of resonances follow from unitarity alone,
without regard to the dynamical mechanism that produces the
resonance. Consequently, as we shall see below, the effect of
resonance-induced FSIs [Fig. 2(a)] can be described in a
model-independent manner in terms of the mass and width of the
nearby resonances.

\begin{figure}[ht]
\vspace{-0.5cm}
  \centerline{\psfig{figure=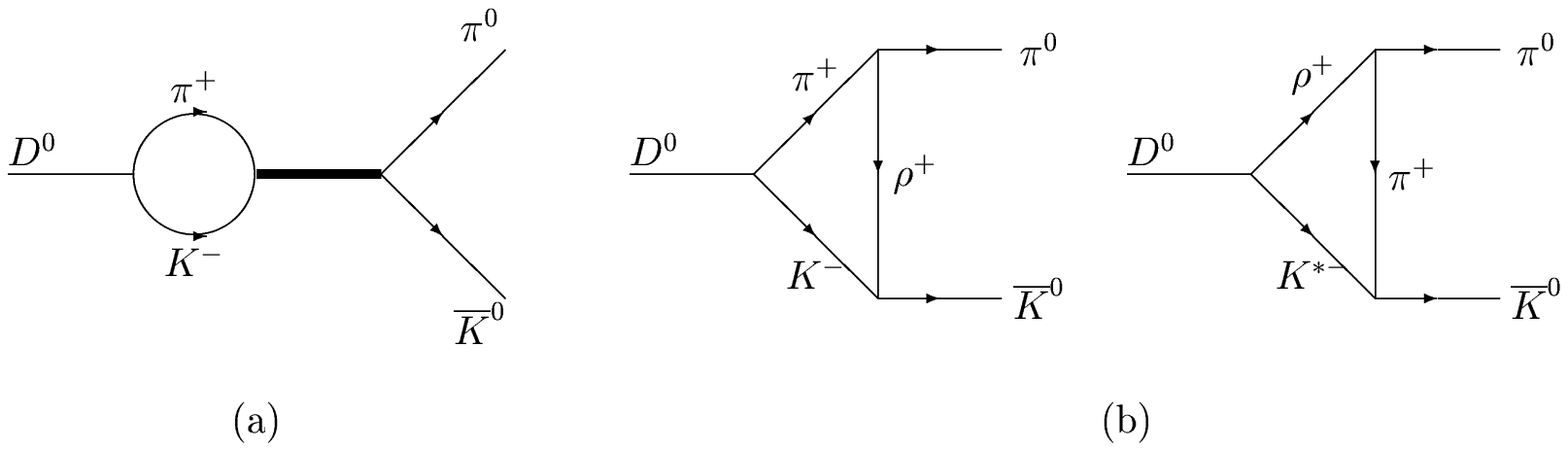,width=16cm}}
\vspace{-17cm}
    \caption[]{\small Manifestation of Fig. 1(a) as the
    long-distance $s$- and $t$-channel contributions to the
    $W$-exchange amplitude in $D^0\to \ov K^0\pi^0$. The thick
    line in (a) represents a resonance.}
\end{figure}

\subsection{Formalism}

In the presence of resonances, the decay amplitude of the charmed
meson $D$ decaying into two mesons $M_1M_2$ is modified by
rescattering through a multiplet of resonances
\cite{Weinberg}\footnote{The same expression for Eq. (\ref{FSI})
is also given in \cite{Buccella} except that the phase in Eq.
(3.3) of \cite{Buccella} is too small by a factor of 2.}
 \be \label{FSI}
 A(D\to M_iM_j)_{\rm resonant-FSI} &=& A(D\to M_iM_j) \non \\
 &-& i{\Gamma\over
E-m_R+i\Gamma/2}\sum_r c^{(r)}_{ij}\sum_{kl} c^{(r)*}_{kl}A(D\to
M_kM_l),
 \en
where the summation runs over various mass degenerated resonances
with the same values of the resonance mass $m_R$ and width
$\Gamma$, and $c^{(r)}_{ij}$ are the normalized coupling constants
of $M_iM_j$ with the resonance $r$, obeying the orthonormal
relations
 \be
\sum_{ij}c_{ij}^{(r)}c_{ij}^{(s)*}=\delta_{rs}, \qquad
\sum_{ij}|c_{ij}^{(r)}|^2=1.
 \en
The presence of a resonance shows itself in a characteristic
behavior of phase shifts near the resonance. For each individual
resonant state $r$, there is an eigenstate of $A(D\to M_iM_j)$
with eigenvalue \cite{Weinberg}
 \be \label{phase}
e^{2i\delta_r}=1-i\,{\Gamma\over m_D-m_R+i\Gamma/2},
 \en
or
 \be
  \tan\delta_r={\Gamma\over 2(m_D-m_R)}
 \en
in the rest frame of the charmed meson. Therefore,
resonance-induced coupled-channel effects are amenable technically
in terms of the physical resonances.

To illustrate the effect of FSIs in the resonance formation,
consider the decays $D^0\to \ov K_i P_j$ as an example. The only
nearby $0^+$ scalar resonance with the $(s\bar d)$ quark content
in the charm mass region is $r=K_0^*(1950)$ and the states $\ov
K_i P_j$ are $K^-\pi^+,\ov K^0\pi^0,\ov K^0\eta,\ov K^0\eta'$. The
quark-diagram amplitudes for $D^0\to K^-\pi^+,~\ov K^0\pi^0,~\ov
K^0\eta_{q}$ and $\ov K^0\eta_s$ are shown in Table I. It is
convenient to decompose $D\to\ov K\pi$ amplitudes into their
isospin amplitudes (see Eq. (\ref{isospin}) and Table I):
 \be \label{qds}
 A(D^0\to(\ov
K\pi)_{3/2}) = {1\over\sqrt{3}}(\t+\e),   \qquad A(D^0\to(\ov
K\pi)_{1/2}) = {1\over\sqrt{6}}(2\t-\c+3\e),
 \en
where the subscripts 1/2 and 3/2 denote the isospin of the $\ov K
\pi$ system. Consider the $D$-type coupling for the strong
interaction $P_1P_2\to P'$ ($P'$: scalar meson), namely
$\kappa{\rm Tr}\left(P'\{P_1,\,P_2\}\right)$ with $\kappa$ being a
flavor-symmetric strong coupling \cite{Zen}. Noting that $(\ov
K\pi)_{3/2}$ does not couple to $(\ov K\pi)_{1/2}$, $\ov
K^0\eta_q$, and $\ov K^0\eta_s$ via FSIs, the matrix $c^2$ arising
from two $D$-type couplings in the $I={1\over 2}$ sector has the
form:
 \be
 c^2 \propto \kappa^2\left(\matrix{ {3\over 2} &
{\sqrt{3}\over 2} & {\sqrt{3}\over\sqrt{2}} \cr   {\sqrt{3}\over
2} & {1\over 2} & {1\over \sqrt{2}} \cr    {\sqrt{3}\over\sqrt{2}}
& {1\over\sqrt{2}} & 1    \cr} \right)
 \en
in the basis of $(\ov K\pi)_{1/2},~\ov K^0\eta_{q},~\ov
K^0\eta_s$. Since $\sum|c_{ij}|^2=3\kappa^2$, it follows that the
normalized matrix $c^2$ reads
 \be \label{hatc}
 c^2\equiv \hat c=\left(\matrix{ {1\over
2} & {1\over 2\sqrt{3}} & {1\over \sqrt{6}}  \cr {1\over
2\sqrt{3}} & {1\over 6} & {1\over 3\sqrt{2}} \cr {1\over\sqrt{6}}
& {1\over 3\sqrt{2}} & {1\over 3} \cr}\right).
 \en
Then it is easily seen that
 \be
 A(D^0 &\to& \ov K^0\eta_s)= \e=e+(e^{2i\delta_r}-1) \cr & \times &
\left[{1\over\sqrt{6}}A(D^0\to (\ov K\pi)_{1/2})+{1\over
3\sqrt{2}}A(D^0\to \ov K^0\eta_{q})+{1\over 3}A(D^0\to\ov
K^0\eta_s)\right],
 \en
and hence
 \be \label{E}
 \e=e+(e^{2i\delta_r}-1)\left(e+{t\over 3}\right),
 \en
where we have used $t,\,c,\,e,\,a$ to denote the corresponding
reduced amplitudes $\t,\,\c,\,\e,\,\a$ before resonant FSIs.
Likewise, it is straightforward to show that $\t=t$ and $\c=c$.
Therefore, resonance-induced FSIs amount to modifying the
$W$-exchange amplitude and leaving the other quark-diagram
amplitudes $\t$ and $\c$ intact. We thus see that even if the
short-distance $W$-exchange vanishes (i.e. $e=0$), as commonly
asserted, a long-distance $W$-exchange contribution still can be
induced from the tree amplitude $\t$ via FSIs in resonance
formation.\footnote{An expression similar to  Eq. (\ref{E}),
 \be
 \e=e+(\cos\delta e^{i\delta}-1)\left(e+{\t\over 3}\right), \non
 \en
was first obtained by Zenczykowski \cite{Zen} by applying the
strong reaction matrix $K_0$ together with the unitarity
constraint of the $S$ matrix to study the effects of
resonance-induced FSIs. However, since
$(e^{2i\delta_r}-1)=2(\cos\delta_r e^{i\delta_r}-1)$, it is clear
that the contribution of resonant FSIs to the $W$-exchange
amplitude as given above is too small by a factor of 2. This has
been corrected in \cite{CT01}.}

Likewise, from Cabibbo-allowed decays $D_s^+\to\ov K^0
K^+,~\pi^+\eta_q$ and $\pi^+\eta_s$  one can show that the
$W$-annihilation amplitude after resonant FSIs reads
 \be \label{A}
 \a=a+(e^{2i\delta_r}-1)\left(a+{\c\over  3}\right).
 \en
Note that, contrary to the $W$-exchange case, the long-distance
$W$-annihilation amplitude is induced from the color-suppressed
internal $W$-emission.

As for the $W$-exchange graph in $D\to \ov K^*\pi$ and $\ov K\rho$
decays, we consider a $0^-$ resonance $P'$ which couples to $VP$
and $PV$ states with the $F$-type coupling, $\kappa'{\rm
Tr}(P'[V,P])$. Proceeding as before, the $6\times 6$ normalized
coupling matrix reads
 \be
 c^2={1\over 2}\left( \begin{array}{r|r}
 \hat c & -\hat c \\  \hline
 -\hat c & \hat c \\ \end{array} \right)
 \en
in the basis of $(\ov K^*\pi)_{1/2}$, $\ov K^{*0}\eta_q$, $\phi\ov
K^0$, $(\ov K\rho)_{1/2}$, $\ov K^0\omega$ and $\eta_s\ov K^{*0}$,
where $\hat c$ is the matrix given by Eq. (\ref{hatc}) and
  \be
&& A(D^0\to(\ov K^*\pi)_{3/2}) = {1\over\sqrt{3}}(\t_V+\e_P),
\qquad A(D^0\to(\ov K^*\pi)_{1/2}) =
{1\over\sqrt{6}}(2\t_V-\c_P+3\e_P),
\non \\
&& A(D^0\to(\ov K\rho)_{3/2}) = {1\over\sqrt{3}}(\t_P+\e_V),
 \qquad A(D^0\to(\ov K\rho)_{1/2}) =
 {1\over\sqrt{6}}(2\t_P-\c_V+3\e_V).
 \en
It is straightforward to show that $\t_{P,V}$ and $\c_{P,V}$ are
not affected by resonant FSIs and
 \be \label{EPV1}
 \e_P &=& e_P+{1\over 2}(e^{2i\delta_r}-1)\left[e_P-e_V+{1\over
3}(\t_V-\t_P)\right], \non \\
 \e_V &=& e_V+{1\over 2}(e^{2i\delta_r}-1)\left[e_V-e_P+{1\over
3}(\t_P-\t_V)\right],
 \en
or
 \be \label{EPV2}
 \e_P+\e_V &=& e_P+e_V, \non \\
 \e_P-\e_V &=& e_P-e_V+(e^{2i\delta_r}-1)\left(e_P-e_V-{1\over
 3}(\t_P-\t_V)\right).
 \en
Note that the terms in square brackets in Eq. (\ref{EPV1}) have
opposite signs for $\e_P$ and $\e_V$ owing to the
charge-conjugation of the strong coupling.\footnote{The expression
of Eq. (\ref{EPV1}) or (\ref{EPV2}) differs from the results
obtained in \cite{Zen,CT99} for $\e_P$ and $\e_V$.} We will come
back to this point later.

As for the $W$-annihilation amplitudes $\a_P$ and $\a_V$, a direct
analysis of resonant FSIs in Cabibbo-allowed decays
$D_s^+\to\rho\pi,\rho^+\eta(\eta')$, $\omega\pi^+,\phi\pi^+,\ov
K^* K$ shows that the reduced amplitudes $\t_{P,V}$ and
$\a_P+\a_V$ are not affected by FSIs in resonance formation
 \be
 \t_P=t_P, \quad\quad \t_V=t_V, \quad\quad \a_P+\a_V=a_P+a_V.
 \en
The relevant normalized coupling matrix in the basis of
$(\rho\pi)_1$, $\ov K^{*0}K^+$ and $K^{*+}\ov K^0$ is given by
 \be
 c^2=\left( \begin{array}{r r r}
  {2\over 3} & {1\over 3} & -{1\over 3} \\
    {1\over 3} & {1\over 6} & -{1\over 6} \\
    -{1\over 3} & -{1\over 6} & {1\over 6} \\
    \end{array} \right),
  \en
 where
 \be
 A(D_s^+\to(\rho\pi)_1)&=&{1\over\sqrt{2}}[A(D_s^+\to\rho^0\pi^+)-A(D_s^+\to\rho^+\pi^0)]
 =\a_P-\a_V,  \non \\
 A(D_s^+\to(\rho\pi)_2)&=&{1\over\sqrt{2}}[A(D_s^+\to\rho^0\pi^+)+A(D_s^+\to\rho^+\pi^0)]
 =0.
 \en
It follows that
 \be \label{APV}
 \a_P+\a_V &=& a_P+a_V, \non \\
 \a_P-\a_V &=& a_P-a_V+(e^{2i\delta_r}-1)\left(a_P-a_V+{1\over
 3}(\c_P-\c_V)\right),
 \en
and $\c_{P,V}$ are not affected. This result was first obtained by
Zenczykowski \cite{Zen}. Note the similarity between the
expressions of (\ref{APV}) and (\ref{EPV2}).

\subsection{Phenomenological implications}
\subsubsection{Resonance-induced weak annihilations}
Eqs. (\ref{E}), (\ref{A}), (\ref{EPV2}) and (\ref{APV}) are the
main results for weak annihilation amplitudes induced from FSIs in
resonance formation. We see that even in the absence of the
short-distance weak annihilation, a long-distance weak
annihilation  can be induced via resonant FSIs. For
parity-violating $D\to PP$ decays, there is a $J^P=0^+$ resonance
$K^*_0(1950)$ in the $s\bar d$ quark content with mass $1945\pm
10\pm 20$ MeV and width $201\pm 34\pm 79$ MeV \cite{PDG}. Assuming
$e=0$ in Eq. (\ref{E}), we obtain
 \be
 \e=1.43\times 10^{-6}\,{\rm exp}(i143^\circ)\,{\rm GeV},
 \en
which is close to the ``experimental" value $\e=(1.60\pm
0.13)\times 10^{-6}\,{\rm exp}(i114^\circ)$ GeV [cf. Eq.
(\ref{PP})]. Presumably, a non-vanishing short-distance $e$ will
bring the phase of $\e$ in agreement with experiment.
Resonance-induced coupled-channel effects are governed by the
width and mass of nearby resonances, which are unfortunately not
well determined. For example, a reanalysis in a $K$-matrix
formalism \cite{Anisovich} quotes $m_R=1820\pm 40$ MeV and
$\Gamma=250\pm 100$ MeV for the same resonance. This leads to
$\e=1.67\times 10^{-6}{\rm exp}(-i158^\circ)\,{\rm GeV}$.
Therefore, we conclude that one needs a $0^+$ resonance heavier
than the charmed meson. Contributions from more distance resonance
at 1430 MeV are smaller (see also \cite{Gronau,Fajfer}).

Since a nearby $0^+$ resonance $a_0$ in the charm mass region has
not been observed, we will not make an estimate of the
resonance-induced $W$-annihilation amplitude in Cabibbo-allowed
$D\to PP$ decay. In the factorization approach, it is expected
that $\e/\a\sim a_2/a_1$ and hence $|\a|>|\e|$.  Nevertheless, it
is clear from Eq. (\ref{A}) that the long-distance
$W$-annihilation is slightly smaller than the $W$-exchange one
because the former is induced from the color-suppressed amplitude
$\c$ and this is consistent with Eq. (\ref{E}).

To estimate the resonance-induced $W$-annihilation amplitude
$\a_P-\a_V$, we employ $\pi(1800)$ as the appropriate $0^-$
resonance with $m_R=1801\pm13$ MeV and $\Gamma=210\pm 15$ MeV
\cite{PDG}. Assuming $a_P-a_V=0$, we find from Eq. (\ref{APV})
that
 \be \label{aP-aVtheor}
|\a_P-\a_V|=2.4\times 10^{-7},
 \en
which is in agreement with the experimental value given by Eq.
(\ref{aP-aV}). Note that if we set $a_P=a_V=0$, this will lead to
$\a_V=-\a_P$ which will imply a vanishing $D_s^+\to\pi^+\omega$,
in contradiction to the experimental observation.

\subsubsection{Hadronic $D$ decays to $\eta$ or $\eta'$}
Weak annihilation effects are crucial for some two-body decays
involving one single isospin component, e.g. the final state
containing an $\eta$ and $\eta'$. This has been discussed in
detail in \cite{CT99,CT01,Fajfer}. To see this, we consider the
Cabibbo-allowed decays $D^0\to\ov K^0(\eta,\eta')$ and
Cabibbo-suppressed modes $D^+\to\pi^+(\eta,\eta')$. In realistic
calculations we use the $\eta\!-\!\eta'$ mixing angle
$\theta=-15.4^\circ$, but it suffices for our purposes to use
$\theta=-19.5^\circ$ to discuss the essence of physics. The
quark-graph amplitudes then read
 \be
 A(D^0\to\ov K^0\eta)=\c/\sqrt{3}, && \qquad A(D^0\to\ov
 K^0\eta')=(\c+3\e)/\sqrt{6}, \non \\
 A(D^+\to \pi^+\eta)={V_{cd}\over
V_{cs}}{1\over\sqrt{3}}(\t+2\c+2\a), && \qquad A(D^+\to
\pi^+\eta')={V_{cd}\over V_{cs}}{1\over\sqrt{6}}(\t-\c+2\a).
 \en
Since $\e$ is comparable to the color-suppressed $\c$ in
magnitude, the decay $D^0\to \ov K^0\eta'$ is largely enhanced by
$W$-exchange. We see from Table VI that its branching ratio is
enhanced by resonance-induced FSIs by almost one order of
magnitude, whereas $D^0\to\ov K^0\eta$ remains essentially
unaffected. Therefore, we conclude that it is the $W$-exchange
effect that accounts for the bulk of $\B(D^0\to\ov K^0\eta')$ and
explains why $\ov K^0\eta'>\ov K^0\eta$.

As for $D^+\to\pi^+(\eta,\eta')$ decays, it is clear from Eq.
(\ref{PP}) that the interference is constructive between $2\a$ and
$\t+2\c$ and destructive between $2\a$ and $\t-\c$. Hence, the
presence of $W$-annihilation is crucial for understanding the data
of $D^+\to\pi^+\eta$ (see Table VI).

\begin{table}[ht]
\caption{Branching ratios (in units of \%) of the charmed meson
decays to an $\eta$ or $\eta'$.}
\begin{center}
\begin{tabular}{ l c c c }
Decay & Without weak annihilation & With weak annihilation & Expt.
\cite{PDG}
\\ \hline
 $D^0\to\ov K^0\eta$ & 0.64 & 0.66  & $0.70\pm 0.10$ \\
 $D^0\to\ov K^0\eta'$ & 0.31 & 1.76 & $1.71\pm 0.26$\\
 $D^+\to\pi^+\eta$ & 0.09 & 0.37 & $0.30\pm 0.06$ \\
 $D^+\to\pi^+\eta'$  & 0.27 & 0.39 & $0.50\pm 0.10$  \\
\end{tabular}
\end{center}
\end{table}

\subsection{Difficulties with weak annihilation amplitudes $\e_{P,V}$ and $\a_{P,V}$}
For the $W$-exchange amplitude in $D\to VP$ decays, at first sight
it appears that its sign flip from $\e_P$ to $\e_V$ can be
naturally explained since the $0^-$ resonance is expected to have
equal and opposite couplings to $K^{*-}\pi^+$ and $K^-\rho^+$, as
shown in Eq. (\ref{EPV1}). It is obvious from Eq. (\ref{EPV2})
that $\e_P=-\e_V$ in the absence of short-distance $W$-exchange
contributions. However, a further study shows some problems.
First, a possible candidate of  the $0^-$ resonance near $m_D$ is
the $K(1830)$ with mass $\sim$ 1830 MeV and width $\sim$ 250 MeV
\cite{PDG}, but only the $K\phi$ decay mode has been reported.
Second, due to the $F$-type coupling of the resonance with $VP$
states, the resonance contribution from the leading tree amplitude
$\t$ is proportional to the difference $\t_P-\t_V$, which is
small. Assuming $e_P=e_V=0$ for the moment, we obtain from Eqs.
(\ref{VP}), (\ref{PV}) and (\ref{EPV1}) that
 \be
 \e_P=-\e_V=1.6\times 10^{-7}{\rm exp}(i17^\circ)
 \en
from the resonance $K(1830)$. Evidently, $\e_P$ and $\e_V$ are
{\it not} governed by resonant FSIs, contrary to the original
conjecture advocated in \cite{Rosner}. In order to explain the
sign flip of the $W$-exchange amplitude, it appears that the
(short-distance) $W$-exchange amplitudes $e_P$ and $e_V$ have to
be sizable in magnitude and opposite in signs or the $0^-$
resonance couples strongly to one of the $K^{*-}\pi^+$ and
$K^-\rho^+$ states, or the long-distance $t$-channel effect
analogous to Fig. 2(b), which has been ignored thus far, gives the
dominant contributions to $\e_P$ and $\e_V$ with opposite signs.
In any of these cases, it is not clear what is the underlying
physics. Therefore, the sign flip of the $W$-exchange amplitude in
Cabibbo-allowed $VP$ decays remains unexplained.

We also face some difficulties for understanding the
$W$-annihilation amplitudes $\a_P$ and $\a_V$ in $D\to VP$ decays.
This is because naively one will expect $D_s^+\to \pi^+\omega$ is
suppressed relative to $D_s^+\to \pi^+\rho^0$. The argument goes
as follows. The direct $W$-annihilation contributions via $c\bar
s\to W\to u\bar d$ are not allowed in
$D_s^+\to\pi^+\omega,\,\rho^+\eta,\,\rho^+\eta'$ decays since the
$(u\bar d)$ has zero total angular momentum and hence it has the
quantum number of $\pi^+$. Therefore, $G(u\bar d)=-$ and the final
states should have odd $G$-parity. Since $G$-parity is even for
$\omega\pi^+$ and odd for $\pi^+\rho^0$, it is clear that the
former does not receive direct $W$-exchange contribution. Can one
induce $D_s^+\to\pi^+\omega$ from resonant FSIs ? The answer is no
because the $J=0,~I=1$ meson resonance made from a quark-antiquark
pair $u\bar d$ has odd $G$ parity. As stressed in \cite{Lipkin},
the even--$G$ state $\pi^+\omega$ (also $\rho\eta$ and
$\rho\eta'$) does not couple to any single meson resonances, nor
to the state produced by the $W$-annihilation diagram with no
gluons emitted by the initial state before annihilation. This is
indeed consistent with Eq. (\ref{APV}) which states that, contrary
to $\a_P-\a_V$, $\a_P+\a_V$ does not receive any $q\bar q'$
resonance contributions. Therefore, it will be expected that
$D_s^+\to\pi^+\omega$ is prohibited (or $\a_V\approx -\a_P$),
whereas $D_s^+\to\pi^+\rho^0$ receives both factorizable and
resonance-induced $W$-annihilation contributions. Experimentally,
it is the other way around: $\B(D_s^+\to\pi^+\omega)=(2.8\pm
1.1)\times 10^{-3}$ \cite{omegapi} and
$\B(D_s^+\to\pi^+\rho^0)\sim 6\times 10^{-4}$ \cite{E791}. Hence,
it is important to understand the origin of the $W$-annihilation
contribution. As noted in passing, Eqs. (\ref{aP+aV}) and
(\ref{aP-aV}) suggest that the phase difference between $\a_P$ and
$\a_V$ is less than $90^\circ$ and that the magnitude of $\a_P$ or
$\a_V$ is smaller than $\e_P$ and $\e_V$, contrary to the $PP$
case. The suppression of $W$-annihilation is expected because of
the $G$-parity constraint. Since $W$-annihilation occurs only in
the $D_s^+$ system for Cabibbo-allowed decays, its suppression may
help to explain why $\tau(D_s^+)>\tau(D^0)$ at the two-body or
quasi-two-body decay level.

\section{Color-suppressed Internal $W$-emission amplitude}
The relative phase between $\t$ and $\c$ indicates some
final-state interactions responsible for this. Fig. 1(b) shows
that final-state rescattering via quark exchange has the same
topology as the color-suppressed internal $W$-emission amplitude.
At the hadron level, Fig. 1(b) is manifested as FSIs with one
particle exchange in the $t$ channel \cite{Zou,Dai}, see Fig. 3.
Note that although Fig. 3 is very similar to Fig. 2(b), the
exchanged particles here are $K^*$ and $K$ rather than $\rho$ and
$\pi$. Another approach is based on the Regge pole model
\cite{Dai}. Admittedly, the estimate of Fig. 3 is subject to some
theoretical uncertainties as discussed before, e.g. the off-shell
form factor appearing in the chiral loop calculation is unknown.
Nevertheless, we still can learn something useful. For example, Li
and Zou \cite{Zou} have calculated rescattering FSIs for
$D^+\to\ov K^{*0}\pi^+$ and $D^+\to\ov K^0\rho^+$ via single pion
exchange. The absorptive part of the chiral loop diagram in analog
to Fig. 3 gives the imaginary contribution to the FSI amplitude.
It was found the imaginary part is negative for $\c_P$ and $\c_V$
(see Table II of \cite{Zou}). This lends a support to the solution
set (i) in Eq. (\ref{PV}). We will leave a detailed study of this
inelastic FSI effect to a separate publication.

\begin{figure}[ht]
\vspace{-1cm}
  \centerline{\psfig{figure=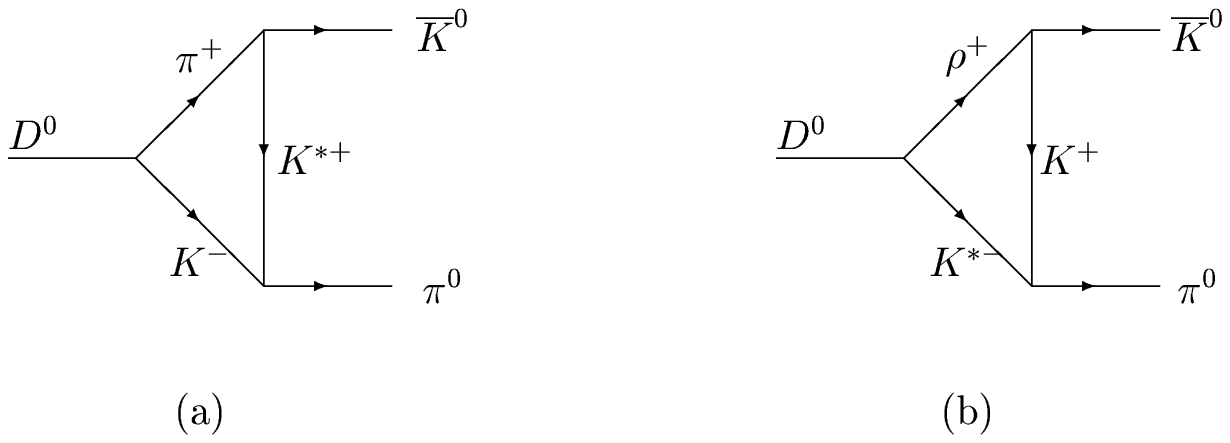,width=16cm}}
\vspace{-17cm}
    \caption[]{\small Manifestation of Fig. 1(b) as the long-distance
    $t$-channel contributions to the
    color-suppressed internal $W$-emission amplitude in $D^0\to \ov K^0\pi^0$.}
\end{figure}

It is often argued that the relative phase between $\t$ and $\c$
can be understood in the factorization approach as arising from
the rescattering through modifications of the phases of isospin
amplitudes \cite{Lipkin80,Neubert,Rosner}. First, one identifies
the isospin amplitudes with the factorizable amplitudes before the
final-state phases are turned on in Eq. (\ref{isospin}). For
example, $A(D^+\to\ov K^0\pi^+)=\sqrt{3}A_{3/2}=T_f+C_f$. This
leads to the isospin amplitudes
 \be
 A_{1/2}={1\over\sqrt{6}}(2T_f-C_f+3E_f), \qquad\qquad
 A_{3/2}={1\over\sqrt{3}}(T_f+C_f),
 \en
where $T_f,C_f,E_f$ are the factorizable tree, color-suppressed,
and $W$-exchange amplitudes, respectively. Second, after
introducing isospin phases to Eq. (\ref{isospin}), it is
straightforward to show that
 \be \label{TC}
 \t+\e &=& (T_f+E_f)e^{i\delta_{1/2}}-{1\over
 3}(T_f+C_f)(e^{i\delta_{1/2}}-e^{i\delta_{3/2}}), \non \\
 \c-\e &=& (C_f-E_f)e^{i\delta_{1/2}}-{2\over
 3}(T_f+C_f)(e^{i\delta_{1/2}}-e^{i\delta_{3/2}}).
 \en
Therefore, the different phases of $\c$ and $\t$ are a consequence
of rescattering. This has an important implication for the
class-II mode $D^0\to\ov K^0\pi^0$. Even if the factorizable $C_f$
and $E_f$ amplitudes are small, the weak decay $D^0\to K^-\pi^+$
followed by the inelastic rescattering $K^-\pi^+\to\ov K^0\pi^0$
can raise $\B(D^0\to\ov K^0\pi^0)$ dramatically and lower
$\B(D^0\to K^-\pi^+)$ slightly.

Of course, the above picture for rescattering is too simplified
and it does not offer a genuine explanation of the phases of
quark-graph amplitudes $\c$ and $\t$. First, the decays $\ov
K^0\eta$ and $\ov K^0\eta'$ have only one isospin component and
the above analysis does not give a clue as to why they are not
color suppressed. Second, the isospin phase difference in $D\to
\ov K\rho$ decays is near zero and yet a relative phase of order
$150^\circ$ between $\c$ and $\t$ is found in the diagrammatic
approach [see Eq. (\ref{PV})]. Third, the above isospin analysis
has no power of prediction as the phase difference is unknown from
the outset; even if elastic $\ov K\pi$ scattering is measured at
energies $\sqrt{s}=m_D$, the isospin phases appearing in
(\ref{isospin}) and (\ref{TC}) cannot be identified with the
measured strong phases.\footnote{If there are only a few channels
are open as the case of two-body nonleptonic decays of kaons and
hyperons, the isospin phases there (or decay amplitude phases) are
related to strong-interaction eigenphases. However, when there are
many channels open and some channels coupled, as in $D$ and
especially $B$ decays, the decay phase is no longer the same as
the eigenphase in the $S$-matrix.}

\section{Comparison with $B$ decays}
It is instructive to compare the present study with the $B$
decays. To proceed, we quote some of the results for $\ov B\to
D\pi,D^*\pi$ decays \cite{cheng01}
 \be \label{BDpi}
\ov B\to D\pi:&& \quad {A_{1/2}\over
A_{3/2}}=(1.00\pm0.14)\,e^{i29^\circ}, \qquad  {a_2\over
a_1}=(0.40\sim  0.65)e^{i59^\circ}, \non \\
 \ov B\to D^*\pi:&& \quad {A_{1/2}\over
 A_{3/2}}=(1.05\pm0.10)\,e^{i29^\circ}, \qquad {a_2\over a_1}=(0.30\sim
 0.55)e^{i63^\circ}.
 \en
The relative phase between $a_1$ and $a_2$ is of order $60^\circ$.
The QCD factorization approach \cite{BBNS} implies that
$\delta_{1/2}-\delta_{3/2}={\cal O}(\Lambda_{\rm QCD} /m_Q)$ and
$A_{1/2}/(\sqrt{2}A_{3/2})=1+{\cal O}(\Lambda_{\rm QCD} /m_Q)$. We
see that, except for $D\to \ov K\rho$, the isospin phase
difference indeed decreases from charm (of order $150^\circ$) to
the bottom system and the ratio of isospin amplitudes in $D$
decays shows a sizable departure from the heavy quark limit. The
relative phase of $a_1$ and $a_2$ is crucial for understanding the
destructive interference in the class-III mode $D^+\to \ov
K^0\pi^+$ and the constructive interference in $B^-\to D^0\pi^-$.

Since the color-suppressed mode $B\to J/\psi K$ does not receive
any weak annihilation contribution and the penguin contribution to
this decay is very tiny, the parameter $a_2$ can be directly
determined from experiment with the result $|a_2(B\to J/\psi
K)|=0.26\pm 0.02$ \cite{a1a2}. It is evident that even in $B$
decays, $a_2$ varies from channel to channel.

In obtaining Eq. (\ref{BDpi}) we have neglected the $W$-exchange
amplitudes in $\ov B\to D^{(*)}\pi$ decays. It is generally argued
that weak annihilation is negligible as helicity suppression
should be more effective due to the large energy release in $B$
decays. Owing to the absence of nearby resonances in the $B$ mass
region, the weak annihilation amplitude will not be dominated by
resonance-induced FSIs, contrary to the charm case. Hence, the
formalism developed in Sec. III for $D$ decays is not applicable
to $B$ mesons.

Recently, there are some growing hints that penguin-induced weak
annihilation is important. For example, the theory predictions of
the charmless $B$ decays to $\ov K^*\pi,~\ov K\rho$ and $\ov
K^*\eta$ based on QCD factorization are too small by one order of
magnitude in decay rates \cite{Du}. This implies that the weak
annihilation (denoted by $P_e$ or $P_a$ in the literature) may
play an essential role. Indeed, a recent calculation based on the
hard scattering pQCD approach indicates that the $\ov K^*\pi$
modes are dominated by penguin-induced annihilation diagrams
\cite{Keum}. This is understandable because the usual helicity
suppression argument works only for the weak annihilation diagram
produced from $(V-A)(V-A)$ operators. However, weak annihilations
induced by the $(S-P)(S+P)$ penguin operators are no longer
subject to helicity suppression and hence can be sizable.

It is anticipated that the soft FSI contributions to the
color-suppressed topology $\c$ are dominated by inelastic
rescattering \cite{Donoghue96}. As for the phase of the ratio of
$a_2/a_1$, the rescattering contribution via quark exchange,
$D^+\pi^-\to D^0\pi^0$, to the topology $\c$ in $\ov B^0\to
D^0\pi^0$ has been estimated in \cite{Blok} using $\rho$
trajectory Regge exchange. It was found that the additional
contribution to $D^0\pi^0$ from rescattering is mainly imaginary:
$a_2(D\pi)/a_2(D\pi)_{\rm without~FSIs}=1+0.61{\rm
exp}(73^\circ)$. This analysis suggests that the rescattering
amplitude can bring a large phase to $a_2(D\pi)$ as expected.

\section{Conclusions}

We have presented a study of hadronic charm decays within the
framework of the diagrammatic approach. We draw some conclusions
from the analysis:

\begin{enumerate}
\item Based on SU(3) symmetry, many of the topological quark-graph
amplitudes for Cabibbo-allowed $D$ decays can be extracted from
the data. The ratio of $a_2/a_1$ has the magnitude of order 0.60,
0.40 and 0.83 for $D\to \ov K\pi,\,\ov K^*\pi,\,\ov K\rho$ decays,
respectively,  with a phase of order $150^\circ$. This implies
that nonfactorizable corrections to $\ov K\rho$ are far more
important than $\ov K^*\pi$.

\item Except for the $W$-annihilation topology in $VP$ decays,
the weak annihilation ($W$-exchange or $W$-annihilation) amplitude
has a sizable magnitude comparable to the color-suppressed
internal $W$-emission with a large phase relative to the tree
amplitude. It receives long-distance contributions from nearby
resonance via inelastic final-state interactions from the leading
tree or color-suppressed amplitude. The effects of
resonance-induced FSIs can be described in a model independent
manner and are governed by the mass and decay width of the nearby
resonances.

\item Weak annihilation topologies in $D\to PP$ decays are dominated by
nearby scalar resonances via final-state resacttering. In
contrast, $W$-exchange in $VP$ systems receives little
contributions from resonant final-state interactions.

\item The experimental data indicate that the three decay amplitudes of
$D\to \ov K\rho$ are essential in phase with one another. This
requires that either the $W$-exchange amplitudes in $\ov K\rho$
and $\ov K^*\pi$ have opposite signs or the relative phase between
the tree and color-suppressed amplitudes flips the sign. While the
latter possibility is probably ruled out by the measurement of
$D^0\to \ov K^{*0}\eta$ and by the model calculation of the phase
of the color-suppressed amplitude, the first possibility is
hampered by the observation that dominance of nearby resonances is
not operative for the $W$-exchange contribution in Cabibbo-allowed
$D\to VP$ decays. Therefore, why the sign of the $W$-exchange
amplitude flips in $D\to\ov K^*\pi$ and $\ov K\rho$ decays remains
unexplained.

\item Owing to the $G$-parity constraint, the $W$-annihilation amplitude
$\a_P$ or $\a_V$ in $D\to VP$ decays is suppressed relative to
$W$-exchange, contrary to the $D\to PP$ case where $W$-exchange
and $W$-annihilation are comparable. Since $W$-annihilation occurs
only in the $D_s^+$ system for Cabibbo-allowed decays, this may
help to explain the longer lifetime of $D_s^+$ than $D^0$.

\item Weak annihilation contributions are crucial for
understanding the data of $D^0\to\ov K^0\eta'$ and
$D^+\to\pi^+\eta$.

\item The relative phase between the tree and color-suppressed
amplitudes arises from the final-state rescattering via quark
exchange. This can be evaluated by considering the $t$-channel
chiral-loop effect or by applying the Regge pole method. Much more
work along this line is needed.

\item Some Cabibbo-suppressed modes exhibit huge SU(3)-flavor
symmetry breaking effects. It can be accounted for by the
accumulation of several modest SU(3) violations in individual
quark-graph amplitudes.

\end{enumerate}

\vskip 3.0cm \acknowledgments  We wish to thank C.N. Yang
Institute for Theoretical Physics at SUNY Stony Brook for its
hospitality. This work was supported in part by the National
Science Council of R.O.C. under Grant No. NSC90-2112-M-001-047.



\begin{thebibliography}{99}
\newcommand{\bi}{\bibitem}

\bibitem{Fuk}
M. Fukugita, T. Inami, N. Sakai, and S. Yazaki, \pl {\bf 72B}, 237
(1977); D. Tadi\'c and J. Trampeti\'c, {\sl ibid.} {\bf 114B}, 179
(1982); M. Bauer and B. Stech, {\sl ibid.} {\bf 152B}, 380 (1985).

\bibitem{Buras}
A.J. Buras, J.-M. G\'erard, and R. R\"uckl, \np {\bf B268}, 16
(1986).

\bi{BS} B. Blok and M. Shifman, {\sl Sov. J. Nucl. Phys.} {\bf
45}, 35, 301, 522 (1987).

\bi{Halperin} I. Halperin, \pl {\bf B349}, 548 (1995).

\bi{BBNS} M. Beneke, G. Buchalla, M. Neubert, and C.T. Sachrajda,
\prl {\bf 83}, 1914 (1999); \np {\bf B591}, 313 (2000); {\sl
ibid.} {\bf B606}, 245 (2001).

\bi{Donoghue} J.F. Donoghue, \pr {\bf D33}, 1516 (1986).

\bi{CC87} L.L. Chau and H.Y. Cheng, \pr {\bf D36}, 137 (1987); \pl
{\bf B222}, 285 (1989).

\bi{Chau} L.L. Chau, {\sl Phys. Rep.} {\bf 95}, 1 (1983).

\bi{CC86} L.L. Chau and H.Y. Cheng, \prl {\bf 56}, 1655 (1986).

\bi{Rosner} J.L. Rosner, \pr {\bf D60}, 114026 (1999).

\bi{CC89} L.L. Chau and H.Y. Cheng, \pr {\bf D39}, 2788 (1989);
L.L. Chau, H.Y. Cheng, and T. Huang, \zp {\bf C53}, 413 (1992).

\bi{Li} X.Y. Li and S.F. Tuan, DESY Report No. 83-078
(unpublished); X.Y. Li, X.Q. Li, and P. Wang, {\sl Nuovo Ciemento}
{\bf 100A}, 693 (1988).

\bi{Dighe} A.S. Dighe, M. Gronau, and J.L. Rosner, \prl {\bf 79},
4333 (1997).

\bi{Lipkin80} H.J. Lipkin, \prl {\bf 44}, 710 (1980).

\bi{PDG} Particle Data Group, D.E. Groom {\it et al.,} {\sl Eur.
Phys. J.} {\bf C15}, 1 (2000).

\bi{Chau91} L.L. Chau, H.Y. Cheng, W.K. Sze, H. Yao, and B. Tseng,
\pr {\bf D43}, 2176 (1991).

\bi{Kroll} T. Feldmann and P. Kroll, {\sl Eur. Phys. J.} {\bf C5},
327 (1998); T. Feldmann, P. Kroll, and B. Stech, \pr {\bf D58},
114006 (1998); \pl {\bf B449}, 339 (1999); T. Feldmann and P.
Kroll, hep-ph/0201044.

\bi{CL} F.E. Close and H.J. Lipkin, \pl {\bf B405}, 157 (1997).

\bi{E791} Fermilab E791 Collaboration, E.M. Aitala {\it et al.,}
\prl {\bf 86}, 765 (2001).

\bi{Buccella} F. Buccella, M. Lusignoli, and A. Pugliese, \pl {\bf
B379}, 249 (1996); F. Buccella, M. Lusignoli, G. Miele, A.
Pugliese, and P. Santorelli, \pr {\bf D51}, 3478 (1995).

\bi{Ball} P. Ball, J.-M. Fr\`ere, and M. Tytgat, \pl {\bf B365},
367 (1996).

\bi{CT99} H.Y. Cheng and B. Tseng, \pr {\bf D59}, 014034 (1999).

\bi{CT01} H.Y. Cheng and B. Tseng, {\sl Chin. J. Phys.} {\bf 39},
28 (2001) [hep-ph/0006081].

\bi{BSW} M. Wirbel, B. Stech, and M. Bauer, \zp {\bf C29}, 637
(1985); M. Bauer, B. Stech, and M. Wirbel, {\sl ibid.} {\bf C34},
103 (1987).

\bi{MS} D. Melikhov and B. Stech, \pr {\bf D62}, 014006 (2001).

\bi{CLEO93} CLEO Collaboration, A. Bean {\it et al.,} \pl {\bf
B317}, 647 (1993).

\bi{LCSR} A. Khodjamirian, R. Ruckl, S. Weinzierl, C.W. Winhart,
and O. Yakovlev, \pr {\bf D62}, 114002 (2000).

\bi{Abada} A. Abada {\it et al.,} \np {\bf B619}, 565 (2001).

\bi{cheng01} H.Y. Cheng, hep-ph/0108096, to appear in Phys. Rev.
D.

\bi{Buras96} G. Buchalla, A.J. Buras, and M.E. Lautenbacher, {\sl
Rev. Mod. Phys.} {\bf 68}, 1125 (1996).

\bi{CC94} L.L. Chau and H.Y. Cheng, \pl {\bf B333}, 514 (1994).

\bi{Dai} Y.S. Dai, D.S. Du, X.Q. Li, Z.T. Wei, and B.S. Zou, \pr
{\bf D60}, 014014 (1999).

\bi{Eeg} J.O. Eeg, S. Fajfer, and J. Zupan, \pr {\bf D64}, 034010
(2001).

\bi{Zen} P. \.Zenczykowski, {\sl Acta Phys. Polon.} {\bf B28},
1605 (1997) [hep-ph/9601265].

\bi{Neubert} M. Neubert, \pl {\bf B424}, 152 (1998).

\bi{aaoud} El hassan El aaoud and A.N. Kamal, {\sl Int. J. Mod.
Phys.} {\bf A15}, 4163 (2000).

\bi{Ablikim} M. Ablikim, D.S. Du, and M.Z. Yang, hep-ph/0201168.

\bi{Gronau} M. Gronau, \prl {\bf 83}, 4005 (1999).

\bi{Weinberg} S. Weinberg, {\it The Quantum Theory of Fields,
Volume I} (Cambridge, 1995), Sec. 3.8.


\bi{Anisovich} A.V. Anisovich and A.V. Sarantsev, \pl {\bf B413},
137 (1997).

\bi{Fajfer} S. Fajfer and J. Zupan, {\sl Int. J. Mod. Phys.} {\bf
A14}, 4161 (1999).

\bi{Lipkin} H.J. Lipkin, \pl {\bf B283}, 412 (1992); in {\sl
Proceedings of the 2nd International Conference on $B$ Physics and
$CP$ Violation}, Honolulu, Hawaii, 1997, edited by T.E. Browder
{\it et al.} (World Scientific, Singapore, 1998), p.436.

\bi{omegapi} CLEO Collaboration, R. Balest {\it et al.,} \prl {\bf
79}, 1436 (1997).

\bi{Zou} X.Q. Li and B. Zou, \pl {\bf B399}, 297 (1997).

\bi{a1a2} H.Y. Cheng and K.C. Yang, \pr {\bf D59}, 092004 (1999).

\bi{Du} D.S. Du, H. Gong, J. Sung, D. Yang, and G. Zhu,
hep-ph/0201253.

\bi{Keum} Y.Y. Keum, H.n. Li, A.I. Sanda, hep-ph/0201103.

\bi{Donoghue96} J.F. Donoghue, E. Golowich, A.A. Petrov, and J.M.
Soares, \prl {\bf 77}, 2178 (1996).

\bi{Blok} B. Blok and I. Halperin, \pl {\bf 385}, 324 (1996).



\end{thebibliography}
\end{document}